\documentclass[onecolum]{revtex4-2}
\usepackage{graphicx}% Include figure files
\usepackage{dcolumn}% Align table columns on decimal point
\usepackage{bm,amsmath,amssymb,color}% bold math
\usepackage{multirow}
\usepackage{longtable}
\usepackage{enumerate}
\usepackage{lineno}
\begin{document}

\preprint{APS/123-QED}

\title{Constraints on the $e^{\pm } $ Pair Injection of Pulsar Halos: \\ Implications from the Galactic Diffuse Multi-TeV Gamma-ray Emission }% Force line breaks with \\
\author{Kai Yan}
\affiliation{School of Astronomy and Space Science, Nanjing University, Nanjing 210023, China}

\author{Ruo-Yu Liu}
\email{ryliu@nju.edu.cn}
\affiliation{School of Astronomy and Space Science, Nanjing University, Nanjing 210023, China}
\affiliation{Key laboratory of Modern Astronomy and Astrophysics (Nanjing University), Ministry of Education, Nanjing 210023, China}
%\correspondingauthor{Ruo-Yu Liu}


\date{\today}% It is always \today, today,
%\linenumbers
\begin{abstract}
Diffuse gamma-ray emission (DGE) has been discovered over the Galactic disk in the energy range from sub-GeV to sub-PeV. While it is believed to be dominated by the pionic emission of cosmic ray (CR) hadrons via interactions with interstellar medium, unresolved gamma-ray sources may also be potential contributors. TeV gamma-ray halos around middle-aged pulsars have been proposed as such sources. Their contribution to DGE, however, highly depends on the injection rate of electrons and the injection spectral shape, which are not well determined based on current observations. The measured fluxes of DGE can thus provide constraints on the $e^\pm$ injection of the pulsar halo population in turn. In this paper, we estimate the contribution of pulsar halos to DGE based on the ATNF pulsar samples with taking into account the off-beamed pulsars.  The recent measurement on DGE by Tibet AS$\gamma$ and an early measurement by MILAGRO are used to constrain the pair injection parameters of the pulsar halo population. Our result may be used to distinguish different models for pulsar halos. 
\end{abstract}

\maketitle

%\tableofcontents

\section{\label{sec:Intro}Introduction}

The Galactic diffuse gamma-ray emission (DGE) is the most prominent structure in the gamma-ray sky, which appears as a bright band associated with the Galactic plane. Measurement of DGE over the entire Galactic plane from tens of MeV up to TeV energies has been done by Fermi Large Area Telescopes (Fermi-LAT) \citep{2012Fermi,2020Neronov}.  Ground-based detectors such as MILAGRO \citep{2005MILAGRO} and ARGO-YBJ \citep{2015ARGO} have measured the DGE in a fraction of the Galactic plane due to their limited observable sky and extend the DGE spectrum to several TeV. More recently, the Tibet $\rm AS\gamma$ experiment \citep{2021ASgamma} reported discovery of diffuse gamma-ray emission between 100\,TeV and 1\,PeV in the Galactic disk for the first time. 

The main component of the DGE is believed to be generated by cosmic-ray (CR) hadrons, which are mostly protons, interacting with interstellar medium (ISM). However, due to the limited sensitivity of instruments, contributions from some faint, extended sources may also be counted in the diffuse emission, such as TeV pulsar halos, from sub-TeV band to sub-PeV band \citep{2018Linden, 21Liu, 2022Fang, Vecchiotti22a, Vecchiotti22b}. 

TeV pulsar halos are spatially extended gamma-ray emissions around middle-aged pulsars. Due to proper motions, these middle-aged pulsars have escaped their associated supernova remnants and are traversing the interstellar medium (ISM). Energetic $e^{\pm}$ pairs (hereafter we do not distinguish electrons from positrons, unless otherwise specified) that are accelerated in their pulsar wind nebulae can escape to the surrounding ISM, up-scattering the cosmic microwave background (CMB) as well as infrared radiation field in the Galaxy \citep{2022Lopez-Coto, 22Liu}, forming halo-like gamma-ray sources. TeV halos are firstly discovered at multi-TeV band by the High Altitude Water Cherenkov telescope (HAWC) around two nearby middle-aged pulsars, namely, PSR J0633+1746 (the Geminga pulsar) and PSR B0656+14 (the pulsar in the Monogem ring, also referred to as the Monogem pulsar) \citep{2017Abeysekara}. Recent observations of the Large High Altitude Air Shower Observatory (LHAASO) \citep{2021LHAASO} identified another pulsar halos, LHAASO~J0621+3755, with spectrum extending beyond 100\,TeV. 

The steep TeV gamma-ray surface brightness profiles of pulsar halos measured by HAWC and LHAASO intuitively indicate a suppressed diffusion zone around the pulsars, although the data can be also explained with the standard interstellar diffusion coefficient under certain conditions \citep{Liu19prl, Recchia21}. Regardless of the on-going debate on the particle transport mechanism in pulsar halos \citep{Yan22,Luque22, Bao22}, even with the standard diffusion coefficient, i.e., $D(E)=4\times10^{28}(E_e/{\rm 1GeV})^{1/3}\,\rm cm^2/s$ \citep[e.g.,][]{11Trotta}, high-energy electrons will cool via synchrotron and inverse Compton (IC) radiation before diffusing a distance of $r_{\rm diff}=2\sqrt{D(E)t_{\rm cool}(E)}\simeq 400(E_e/100\,\rm TeV)^{-1/3}\,$pc, where we consider a Galactic magnetic field of $3\mu$G for the synchrotron radiation loss and the CMB radiation field as the target for the IC radiation loss for the cooling timescale of electrons $t_{\rm cool}$. This diffusion distance is comparable to the thickness of the Galactic disk, implying that injected high-energy electrons in pulsar halos will deposit most of their energies in the Galactic disk {\bf (for more details see Appendix~\ref{app:a})}. The detailed amount of deposited energy goes into the TeV band depends on some model parameters such as the injection spectral shape and the efficiency of spindown energy of pulsars being converted to energies of electrons.  

\citet{2018Linden} assumed every young and middle-aged pulsar can power a gamma-ray halo, and simulated a steady-state pulsar population in the Galaxy based on the supernova rate and the distribution of massive stars, pulsars and supernova remnants. By assigning an initial rotation period and magnetic field to each generated pulsar following the study of the observed pulsar population, they showed that the injected electrons from pulsars can dominate the DGE at the TeV band and explain MILAGRO data given appropriate choice of the electron injection spectrum. While the true DGE fraction contributed by pulsar halos highly depends on the properties of electron injections, the observed DGE can be conservatively regarded as an upper limit of the gamma-ray emission from the pulsar halo population, which then can be used to constrain electron injection in pulsar halos. 

In this study, we aim to explore the constraints on the electron injection of pulsar halos based on the DGE measured by AS$\gamma$\citep{2021ASgamma} and MILAGRO\citep{2005MILAGRO}. Different from previous studies, we will employ observed pulsar population based on the Australia Telescope National Facility (ATNF) pulsar catalog \citep{2005Manchester} instead of a simulated pulsar population. This is to reduce the poss deviation of the simulation from reality. In addition, using the observed pulsar sample would allow us to predict the contribution from each realistic pulsar. On the other hand, another difference from \citet{2018Linden} will be the exclusion of relatively young pulsars with age less than 100\,kyr in our calculation. Electrons accelerated in those compact young PWNe may be well confined and have not escaped to ISM. They have been likely resolved by many TeV gamma-ray instruments, and are removed in the DGE analysis by AS$\gamma$ for instance. Exclusion of those relatively young pulsars will make our constraints more conservative.

This rest of the paper is organized as follows. In Section 2, we introduce the method to select the pulsar samples and to calculate their contribution to the DGE. In Section 3, we show the main results of this work. In Section 4, we further discuss some model uncertainties. In Section 5, we present our conclusion.

\section{\label{sec:Sec2}Method}

% In this work, we calculate the gamma-ray fluxes from a population of unresolved pulsar halos. One of the differences with respect to the previous studies is that we use samples of real observed pulsars from the Australia Telescope National Facility (ATNF) pulsar catalog and also take off-beamed pulsars into consideration. In addition, we further obtain constraints on injection parameters. The diffuse gamma-ray fluxes generated by the proton-proton (pp) collisions is calculated following the factorized model in \citet{2018Lipari}.

\subsection{\label{sec:Sec2.1}Sample Selection}
There are more than 3000 pulsars recorded in the ATNF catalog. We select pulsars from the ATNF catalog with the following three conditions:\\
\begin{itemize}
    \item with characteristic age between 100 $\rm kyr$ and 10 $\rm Myr$;
    \item within the region of interests (ROI) for AS$\rm \gamma$ (i.e., $25^{\circ}<l<100^{\circ}$, $\mid b\mid < 5^{\circ}$) and for MILAGRO (i.e., $40^{\circ}<l<100^{\circ}$,$\mid b\mid < 5^{\circ}$);
    \item more than 0.5 degree away from the observed TeV sources in the TeVcat \citep{2008TeVcat}.
\end{itemize}
 The first condition is to exclude the contribution of relatively young pulsars and millisecond pulsars. For those relatively young pulsars, the injected electrons may mainly radiate in their PWNe and do not form pulsar halos. There also exists a population of old pulsars which have been spun up through accretion of matter from a donor star in a close binary system, i.e., the millisecond pulsars. They may have quite complex ambient environments than middle-aged pulsars. Although TeV pulsar halos might also exist around millisecond pulsars\citep{22Hooper}, we ignore their possible contribution in this work to be on the conservative side. The third condition mainly follows the treatment of AS$\gamma$ in their analysis of DGE\citep{2021ASgamma}, where they masked a region of $0.5^\circ$ radius centered at each source recorded in the TeVcat in order to exclude contributions of pulsars therein. Note that many TeV sources had not been discovered yet in the era of MILAGRO, and hence the measurement of MILAGRO on DGE should contain a lot of contributions from bright TeV sources which are excluded in the measurement of AS$\gamma$. For simplicity, however, we still perform the masking procedure when comparing with MILAGRO data, since this would only lead to more conservative constraints.

 Among all the pulsars samples, PSR J1952+3252 is of an extremely high spin-down luminosity of $3.7 \times 10^{36}$ $\rm erg \ s^{\rm -1}$, which is 2 orders higher than that of Geminga pulsar, while its distance from Earth is about 3\,kpc. Regardless of the positive theoretical prediction\citep{2003Bednarek}, no TeV gamma-ray emission has been detected from the pulsar or its associated PWN yet. Actually, the pulsar is still located inside a supernova remnant (i.e., CTB 80) implying that it probably has not reached the middle-age stage. This possibility is corroborated by its fast spin with a rotation period of only 40\,ms. Unless the initial rotation period is far less than this value, its true age may be considerably shorter than its characteristic age (i.e., 107\,kyr). We speculate that the accelerated electrons are still confined inside its SNR-PWN complex, and their IC radiation may be suppressed due to a high magnetic field therein. Therefore, we exclude this specific pulsar from our samples. All the pulsar samples selected in our calculation are listed in Appendix \ref{app:b} (see Table~\ref{tab:pulsar}).

\subsection{\label{sec:Sec2.2}Modeling the Gamma-ray Flux of Pulsar Halos}
For a single pulsar halo, the energy losses of electrons through a combination of synchrotron radiation and IC radiation can be given by
\begin{eqnarray}
\frac{\mathrm{d} E_{\mathrm{e}}}{\mathrm{d} t}=-\frac{4}{3} \sigma_{\mathrm{T}} c\left(\frac{E_{\mathrm{e}}}{m_{\mathrm{e}} c^{2}}\right)^{2}\left[U_{B}+\sum_{i}U_{\mathrm{i}} /\left(1+4 \frac{E_{\mathrm{e}} \epsilon_{\rm i}}{m_{\mathrm{e}}^{2} c^{4}}\right)^{3 / 2}\right]
\label{eq:dEe}
\end{eqnarray}
where $\sigma_{\rm T}$ is the Thompson cross section. The magnetic field strength is taken as 3$\rm \mu G$ and the magnetic energy density $U_{\rm \mathrm{B}}$ is $0.22\,\rm eV \ cm^{-3}$. $U_i$ and $\epsilon_i$ represent the energy density and the typical photon energy of the $i$th component of the interstellar radiation field (ISRF) assuming a black body or a grey body distribution with temperature $T_i$ for their spectra (i.e., $\epsilon_i=2.82kT_i$). The considered target radiation field includes the CMB radiation field ($T_{\rm CMB}$ = 2.73\,K and $U_{\rm CMB}$ = 0.26\,$\rm eV \ cm^{-3}$), a far-infrared radiation field ($T_{\rm FIR}$ = 30\,K and $U_{\rm FIR}$ = 0.3\,$\rm eV \ cm^{-3}$), and a visible light radiation field ($T_{\rm VIS}$ = 5000\,K, $U_{\rm VIS}$ = 0.3\,$\rm eV \ cm^{-3}$). The injected electron spectrum following a power-law with an exponential cutoff, denoted as
\begin{eqnarray}
Q\left(E_{\rm e}, t\right)=N_{\rm 0}\left(t \right) E_{\rm e}^{-s} e^{-E_{\rm e} / E_{\rm \max }}
\label{eq:Qe}
\end{eqnarray}
where $s$ is the spectral index and $E_{\rm max}$ is the cutoff energy. Assuming the spin-down energy loss is totally governed by the dipole radiation, i.e. under the condition of braking index $n=3$, the normalization constant $N_{\rm 0}$ can be determined by
\begin{eqnarray}
\int_{\rm E_{\rm 0}}^{\infty} E_{\rm e} Q\left(E_{\rm e},t\right) d E_{\rm e}=\eta_{\rm e} L_{\rm s,j}\left(t \right)=\eta_{\rm e} \frac{L_{0,j}}{\left(1+t / \tau_{0,j}\right)^{2}}
\label{eq:eta}.
\end{eqnarray}
where $\eta_{\rm e}$ represents the fraction of the pulsar spin-down energy that goes into the electrons which is another important input parameter. Here $L_{\rm s}$ is the spin-down luminosity of the pulsar, $L_{\rm 0}$ is the initial spin-down luminosity, and $\tau_{0}$ is the initial spin-down time scale. The subscript $j$ represents the $j$th pulsar within the ROI, as listed in Table~\ref{tab:pulsar}. The minimum integral energy $E_{\rm 0}$ here is set to be 50\,GeV. The age of the pulsar $t_{\rm age,j}$, characteristic age $\tau_{\rm c,j}$, and initial spin-down time scale $\tau_{\rm 0,j}$ are related by
\begin{eqnarray}
t_{\rm \mathrm{age,j}}= \tau_{\rm c,j}- \tau_{\rm 0,j}= \frac{P_{\rm i}}{2\dot{P_{\rm j}}} \left[1-\left(\frac{P_{\rm 0,j}}{P_{\rm j}}\right)^{2}\right]
\label{eq:tage}
\end{eqnarray}
in which $P_{\rm 0,j}$ is the initial rotation period, $P_{j}$ is the current rotation period, and $\dot{P_{\rm j}}$ is the period derivative.
The electrons differential number density at the present time (i.e., $t=t_{\rm age,j}$) for $j$th pulsar halo is calculated by
\begin{eqnarray}
N_{\rm j}\left(E_{\rm \mathrm{e}} \right)=\int_{\rm 0}^{t_{\rm \mathrm{age,j}}} Q\left(E_{\rm e}, t \right) \mathrm{d}t  \frac{\mathrm{d} E_{\rm \mathrm{g}}}{\mathrm{d} E_{\rm \mathrm{e}}} .
\label{eq:NEe}
\end{eqnarray}
where the relationship between initial injected energy $E_{\rm g}$ and  current energy $E_{\rm e}$ can be obtained from Eq.~(\ref{eq:dEe}).
After obtaining $N_{\rm j}\left(E_{\rm \mathrm{e},t_{\rm \mathrm{age,j}}}\right)$, we calculate the IC and synchrotron luminosity $L_{\rm  j}\left(E_{\rm \gamma}\right)$ of the $j$th pulsar halo according to the semi-analytical method given by \citet{2014Khangulyan} and \citet{2013Fouka} respectively. 

To get the total intensity of unresolved pulsar halos, we sum over the contribution of each single pulsar halo. Note that there should exist many pulsars the lighthouse-like radiation beam of which do not sweep Earth as they spin. As a result, we cannot detect them but they may still inject electrons in the surrounding ISM and form pulsar halos. The fraction of those off-beamed pulsar depend on the size of the beam.  Taking into account of pulsar halos from off-beamed pulsars, we weight the contribution of each single pulsar halo by its beaming fraction. It leads to an average intensity
\begin{eqnarray}
I_{\gamma} = \frac{1}{\Omega}\sum_{\rm j}\frac{L_{\rm  j}\left(E_{\rm \gamma}\right)}{4\pi d_{\rm j}^{2}f_{\rm beam,j}} 
\label{eq:Ntot}
\end{eqnarray}
where $f_{\rm beam,j}$ represents the ratio of the solid angle subtended by the radiation beam (which may be related to the magnetic inclination solid angles) of a pulsar to 4$\pi$. Following the study in Ref.\citep{1998Tauris}, we adopt $f_{\rm beam,j} = 0.011\left[\rm log\left( \tau_{\rm c,j}/100\right) \right]^{2}+0.15$ and $\tau_{\rm c,j}$ in unit of $\rm Myr$. $\Omega$ is the corresponding solid angle of the ROI after masking the known TeV sources, which are 0.219\,sr and 0.177\,sr, respectively, for AS$\gamma$ and MILAGRO.

\subsection{\label{sec:Sec2.3} Influence of Model Parameters}

\begin{figure*}[htbp]
\includegraphics[width=\textwidth]{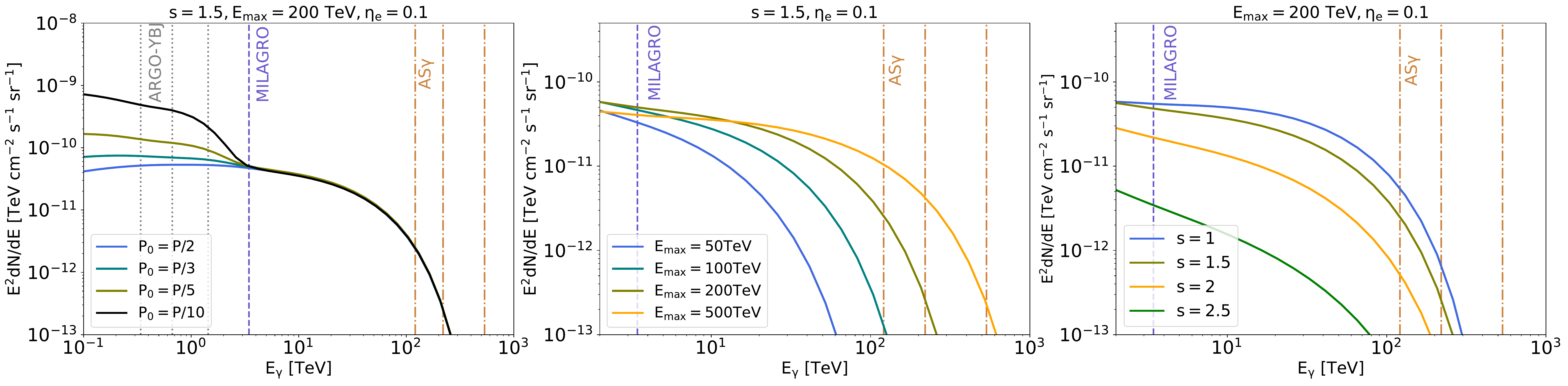}
\caption{Spectral energy distribution (SED) of the population of unresolved pulsar halos within $25^{\circ}<l<100^{\circ}$, $\mid b\mid < 5^{\circ}$. The left panel shows the influence of initial period models on the SED. The middle panel shows the influence of cutoff energy $E_{\rm max}$ on the SED, and the right panel shows that of spectral index $s$. Dashed lines show the observational energies of instruments.}
\label{fig:spec}
\end{figure*}

Before we present our main results, we show the influence of some model parameters which would be helpful to understand the results. The first one is the initial rotation period $P_0$. $P_0$ can be derived based on the assumed spindown history of the pulsar if the true age of the pulsar is known. However, the true ages of most pulsars are unknown and hence we can only make assumptions for $P_0$. Its value has an enormous influence on the early injection history of electrons. The smaller $P_0$ is, the higher initial spindown luminosity the pulsar will get. Therefore, a vast amount of electrons would be injected at early time if a small $P_0$ is assumed. Given the assumed magnetic field and the radiation field, electrons with energy $>10\,$TeV would cool at 100\,kyr. As a result, we would expect electrons injected at early time with energy lower than 10\,TeV are accumulated to the present time and cause a huge flux below $\sim$TeV. As illustrated in the left panel of Fig.~\ref{fig:spec}, we show the influence of initial periods on the resultant gamma-ray spectrum. We see the strong dependency on $P_{\rm 0}$ at $\lesssim$TeV energy, but the spectrum is almost unaffected above several TeV. 

On the other hand, it should be noted that the electromagnetic environment where injected electrons reside could be very different from that of ISM at the early evolutionary stage of the PWN. The magnetic field strength in some young PWNe are found to be much higher than that of the interstellar magnetic field \cite[e.g.,][]{1996Atoyan, 2006Kothes, 2009Gelfand, 2011Hinton, 2012Tang_apj, 2014Torres, 2020Khangulyan, 2022Liang}, electrons injected at early time, which are very likely well confined in the PWN, may cool very efficiently. Also, the expansion of PWN at early time may lead to adiabatic cooling of confined electrons. As a result, these relatively low-energy electrons may not survive at the present time and hence do not produce the GeV-TeV bump as shown in the left panel of Fig.~\ref{fig:spec}. Given such a large uncertainty, we only use the MILAGRO data and the AS$\gamma$ data, which are above several TeV, to constrain the parameters in the following sections and ignore the DGE measured by ARGO-YBJ intentionally. 

%This reflects the cooling history of electrons of different energies. Electrons with energy $E_{\rm e} < 10~ \rm TeV$ cannot be cooled within 100 kyr. Hence, the early injected electrons might be dominant in the corresponding photon flux. A smaller initial period $P_{\rm 0}$ will increase the amount of early injected electrons, thereby increasing the photon fluxes at sub-TeV bands. The ARGO, MILAGRO and AS$\rm \gamma$ data are illustrated with dashed lines in Fig.Since we know little about the initial period of pulsars, the photon fluxes at sub-TeV band have lots of uncertainties, which makes ARGO data not conducive to limiting the injection parameters. In the later contents, we only analyze the constraints of AS$\rm \gamma$ and MILAGRO data on parameters. 

It is straightforward to envisage that the cutoff energy $E_{\rm max}$ and the spectral index $s$ in the injection electron spectrum are important to the resultant diffuse gamma-ray flux. In the middle and right panels of Fig.~\ref{fig:spec}, we compare the resultant gamma-ray spectra with different cutoff energies and injection spectral indexes, respectively, while keeping the total injection luminosity the same. $E_{\rm max}$ and $s$ both affect the gamma-ray spectral shape but do not alter the level of the peak flux significantly. On the contrary, the conversion efficiency $\eta_{\rm e}$, which is proportional to the injection luminosity, can lead to a systematic shift in the amplitude of the gamma-ray flux without changing the spectral shape, which is not illustrated here. As expected, a larger cutoff energy and a harder injection spectrum result in a higher flux at high energy end, and vice versa. Thus, we may expect that the MILAGRO measurement at several TeV and the AS$\gamma$ measurement at sub-PeV would play different roles in constraining injection parameters. 

\begin{figure*}[htbp]
%\hspace{0.9cm} 
\includegraphics[width=0.45\textwidth]{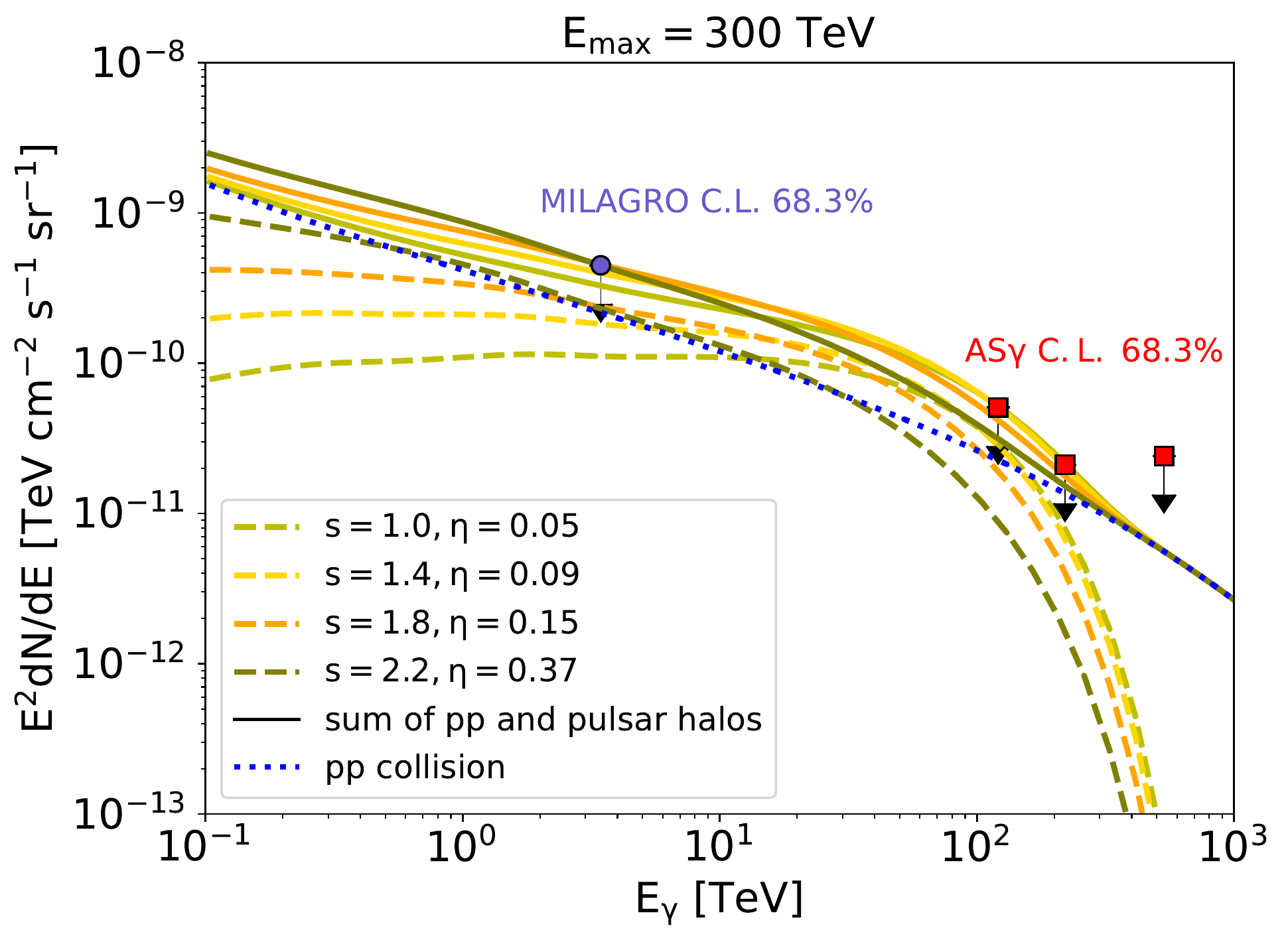}\\
\caption{Expected contribution of unresolved pulsar halos to DGE under different parameter sets. The purple circle marks the 68.3\% upper limit of photon fluxes observed by MILAGRO, and red squares mark 68.3\% upper limits of AS$\rm \gamma$. The blue dotted curve shows the flux of $pp$ collisions by cosmic-ray hadrons. Dashed curves with different colors show the contribution of unresolved pulsar halos with different sets of parameters as labeled in the figure. Solid curves show the total diffuse gamma-ray fluxes of pp collisions and unresolved pulsar halos.}
\label{fig:flux_1sigma}
\end{figure*}

\section{\label{sec:Sec3}Results}

%\begin{figure*}[htbp]
%\hspace{0.9cm} 
%\includegraphics[width=1.0\textwidth]{flux_1sigma.pdf}\\
%\caption{The fit to the DGE from the contribution of unresolved pulsar halos and pp collisions. Purple symbol marks the 68.3\% upper limits of photon fluxes observed by MILAGRO, and red symbols mark 68.3\% upper limits of AS$\rm \gamma$. The blue dotted curve shows the contribution of pp collisions. Dashed curves show the contribution of unresolved pulsar halos with a set of parameters. Solid curves show the total diffuse gamma-ray fluxes of pp collisions and unresolved pulsar halos.}
%\label{fig:flux_1sigma}
%\end{figure*}

In this section, we show the constraints on the electron injection under different sets of model parameters. As mentioned before, the diffuse gamma-ray fluxes measured by AS$\rm \gamma$ and MILAGRO should contain the contribution from both the pionic emission induced by cosmic-ray hadrons and unresolved sources. We adopt the factorized model developed by Ref.~\cite{2018Lipari} to estimate a conservative contribution of the cosmic-ray hadronic component. We then calculate the corresponding IC fluxes under a set of injection parameters from the pulsar halo population, and find out the critical combinations of parameters with which the sum of the hadornic component and the IC component reach the 68.3\% (or 99.7\%) upper limit of any data point measured by either AS$\gamma$ or MILAGRO. The corresponding parameter set is then regarded as the constraints on the electron injection parameters. In Fig.~\ref{fig:flux_1sigma}, we show for instance a few cases that IC emission of pulsar halos and pionic emission of CR hadrons reach the 68.3\% upper limits of DGE observed by AS$\rm \gamma$ or MILAGRO under different combination of parameters. 

%We start from the simplest model setups with considering $E_{\rm max}$ as an independent parameter and take equal values for all pulsars. Furthermore, we take the acceleration abilities of pulsars into consideration and connect $E_{\rm max}$ with the spin-down luminosity of the pulsar. And also, we explore the cases with taking the injection spectrum as a broken power-law. 

\subsection{Baseline Model Setups}

\begin{figure*}[htbp]
%\hspace{0.9cm} 
\centering
\includegraphics[width=0.45\textwidth]{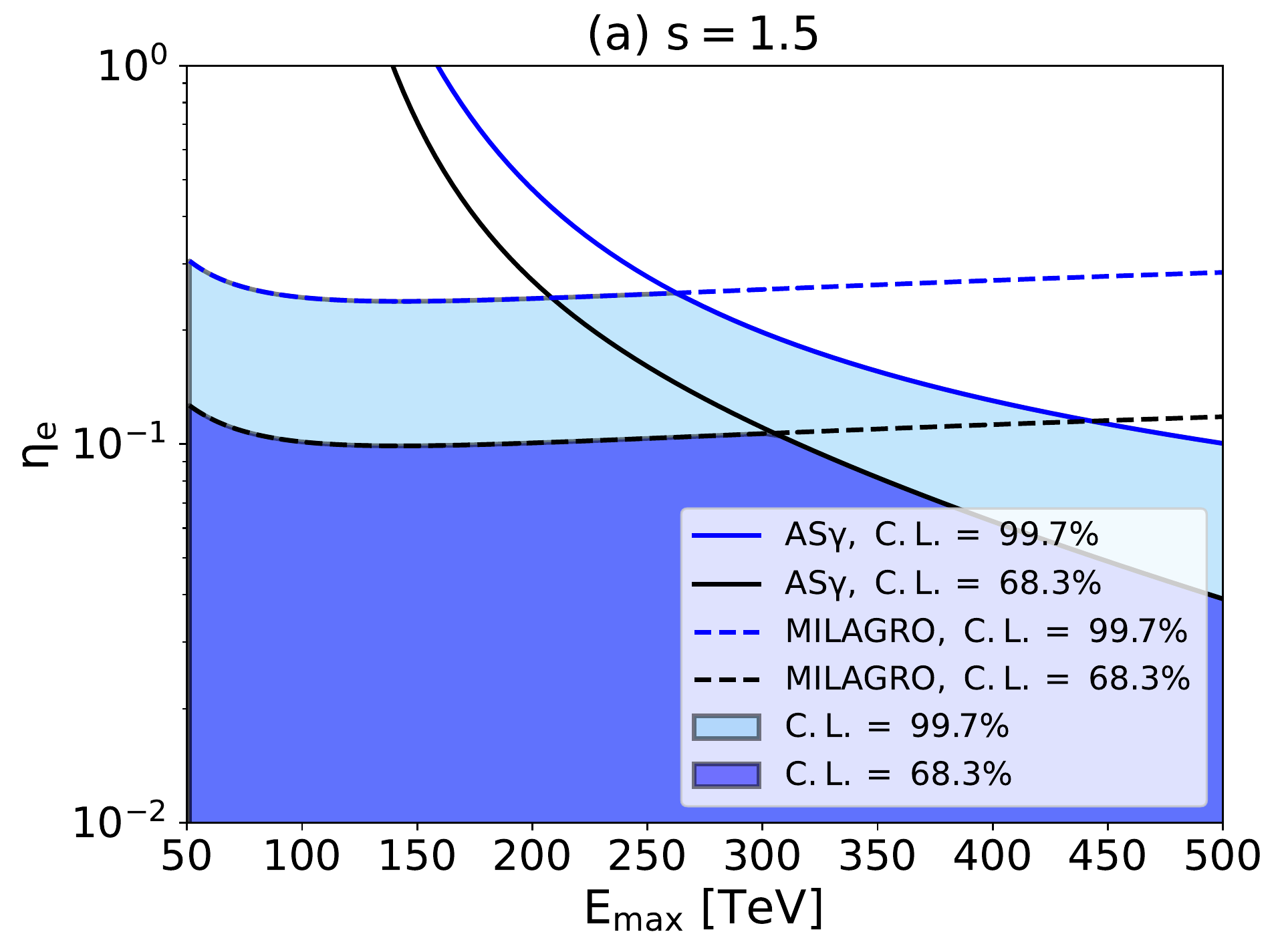}
\includegraphics[width=0.45\textwidth]{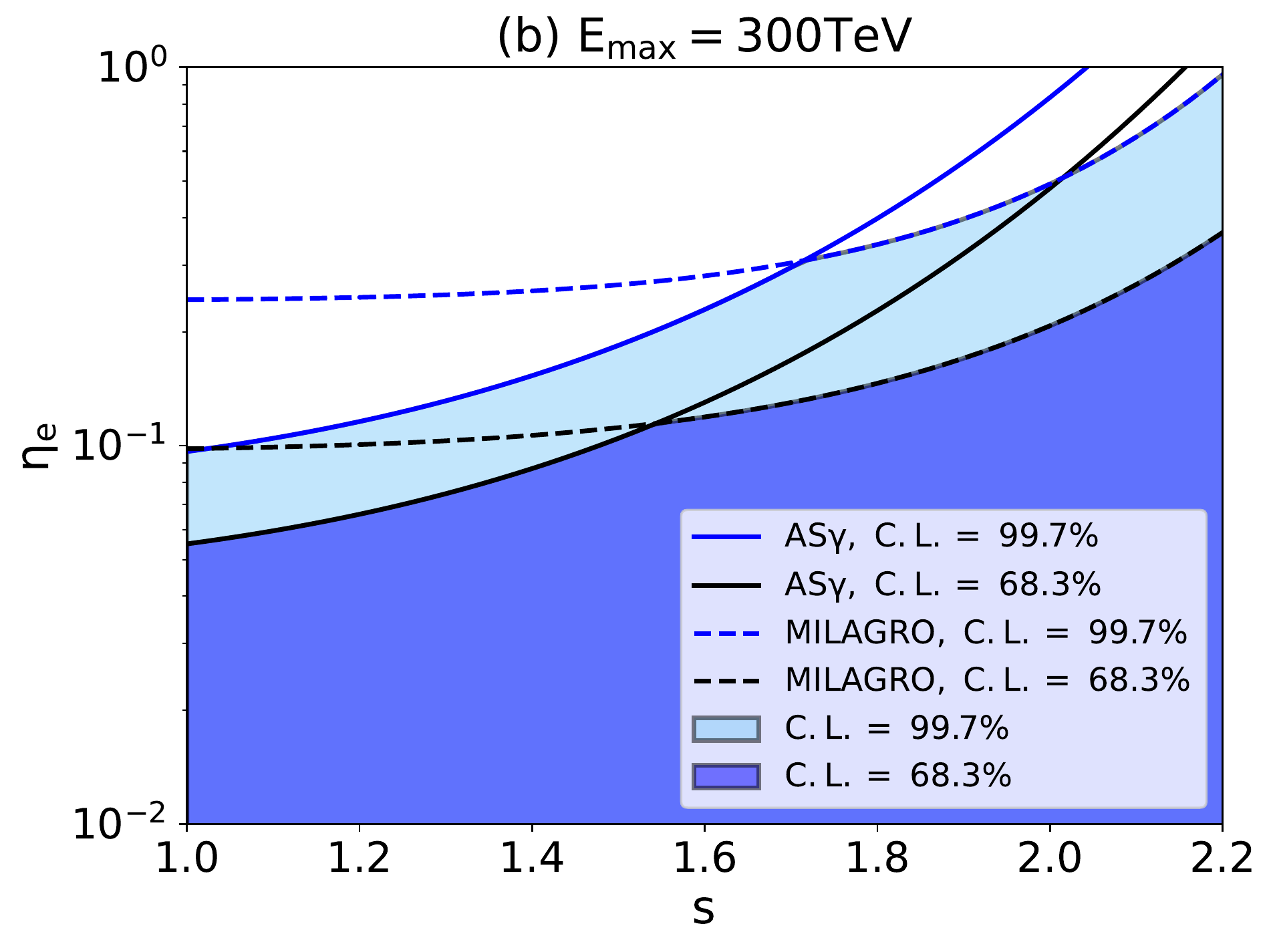}
\caption{(a) two-dimensional constraints between $E_{\rm max}$ and $\eta_{\rm e}$, with $s$ fixed to 1.5 (b) two-dimensional constraints between $s$ and $\eta_{\rm e}$, with $E_{\rm max}$ fixed to $\rm 300~TeV$. Black and blue curves show the boundary of injection parameters that reach 68.3\% confidence level and 99.7\% confidence level of flux, respectively. Solid and dashed curves represent cases for AS$\rm \gamma$ and MILAGRO, respectively. The left panel shows the constraints of combination of $E_{\rm max}$ and $\eta_{\rm e}$, with $s$ fixed. And the right panel shows the constraints of combination of $s$ and $\eta_{\rm e}$, with $E_{\rm max}$ fixed.}
\label{fig:Emax_with}
\end{figure*}

For the sake of simplicity and to be in accordance with the previous modeling setups (e.g. \citep{2017Abeysekara,2021LHAASO,2019DiMauro,2019Tang}), we first consider $E_{\rm max}$ as a common parameter over all pulsar halos. We then explore constraints on the conversion efficiency $\eta_e$ for different $E_{\rm max}$ or spectral index $s$. We present the result in Fig.~\ref{fig:Emax_with}, where the left panel shows the two-dimensional constraints between $E_{\rm max}$ and $\eta_{\rm e}$, with $s$ fixed to 1.5 as the benchmark value. The right panel shows constraints on the combination of $s$ and $\eta_{\rm e}$, with $E_{\rm max}$ fixed to 300\,TeV as the benchmark value. Solid and dashed curves represent constraints from AS$\rm \gamma$ ($25^{\circ}<l<100^{\circ}$, $\mid b\mid < 5^{\circ}$) and MILAGRO ($40^{\circ}<l<100^{\circ}$, $\mid b\mid < 5^{\circ}$), respectively. Black and blue curves show the combination of parameters that reach the 99.7\% and 68.3\% upper limits of photon flux, respectively. Parameters space shaded with cyan and blue are the allowed region corresponding to the 99.7\% and 68.3\% upper limits, respectively. 

We see that the upper limits of $\eta_{e}$ obtained from AS$\rm \gamma$ depend heavily on $E_{\rm max}$, as illustrated in solid lines in Fig~\ref{fig:Emax_with} (a). This is because the injected electron flux at the energy relevant with the energy band of AS$\rm \gamma$, i.e. several hundred of TeV, are very sensitive to the value of $E_{\rm max}$. The resulting gamma-ray fluxes above 100\,TeV have drastic changes when $E_{\rm max}$ increases from 50\,TeV to 500\,TeV, as shown in Fig~\ref{fig:spec}(a). On the other hand, in the energy band of MILAGRO, the photon fluxes and upper limits of $\eta_{e}$ are much less dependent by $E_{\rm max}$ unless $E_{\rm max}$ drops below 20\,TeV. For a fixed $E_{\rm max}$, the upper limit of $\eta_e$ monotonically increases with the injection spectral index $s$ based on either the AS$\rm \gamma$ data or the MILAGRO data, as shown in Fig.~\ref{fig:spec}(b). Combining the two panels, we may conclude that the 68.3\% upper limits of $\eta_{\rm e}$ can be constrained to be $\lesssim 0.1$ for $s < 1.8$. This is a relatively strong parameter restriction, considering 0.1 is a commonly taken value in literature (e.g. \citep{2018Linden}). For $s > 2$, both AS$\rm \gamma$ and MILAGRO data give weak constraints on $\eta_{e}$, since the resulting gamma-ray fluxes peak at lower energy ranges. However, as we mentioned earlier, the DGE data at lower energies does not help because of the large uncertainty in theoretical prediction.

\subsection{$E_{max}$ as a Function of Spin-down Luminosity}

The cutoff energy $E_{\rm max}$ is considered to be the same for all pulsars in the previous section. In reality, $E_{\rm max}$ may vary from pulsar to pulsar, depending on their properties. It has been suggested \cite{1992dejager, 2021Amato, 2022emma} that the maximum achievable energy of electrons in a PWN depends on the maximum potential drop between the pulsar and infinity. Based on the $Hillas$ criterion, the maximum energy can be given by $E_{\rm max}=eB_{\rm TS}R_{\rm TS}$ regardless of the acceleration mechanism, where $B_{\rm TS}$ and $R_{\rm TS}$ are the magnetic field and size of the termination shock respectively. Denoting the ratio of the magnetic energy to the pulsar wind energy by $\eta_{B}$, we have $B_{\rm TS}^{2}/8 \pi=\eta_{B} L_{\rm s}/\left(4 \pi R_{\mathrm{TS}}^{2} c\right)$. The magnetic field at the termination shock $B_{\rm TS}$ can be estimated as $B_{\rm TS}= (2\eta_{\rm B})^{1/2}R_{\rm TS}^{-1}(L_{\rm S}/c)^{1/2}$ then. As a result, 
\begin{equation}
E_{\rm max}=(2\eta_{\rm B})^{1/2}e(L_{\rm s}/c)^{1/2}
\label{eq:etaB}.
\end{equation}
which is expressed as a fraction of $(2\eta_{\rm B})^{1/2}$ of the pulsar potential drop $\Phi_{\rm PSR}=e(L_{\rm s}/c)^{1/2}$. We may find that $\eta_{\rm B} \approx 0.2(E_{\rm max}/100\,{\rm TeV})^2(L_{\rm s}/10^{34} \,\rm erg~s^{-1})^{-1}$ based on Eq.~(\ref{eq:etaB}). Previous studies have shown that $E_{\rm max}\gtrsim 100\,$TeV is needed to model the spectra of the observed pulsar halos (e.g., $E_{\rm max}=150\,$TeV for LHAASO~J0621+3755\citep{2021LHAASO}), which implies $\eta_{\rm B}$ should not be smaller than 0.1 for middle-aged pulsar halos.

\begin{figure*}[htbp]
%\hspace{0.9cm} 
\centering
\includegraphics[width=0.45\textwidth]{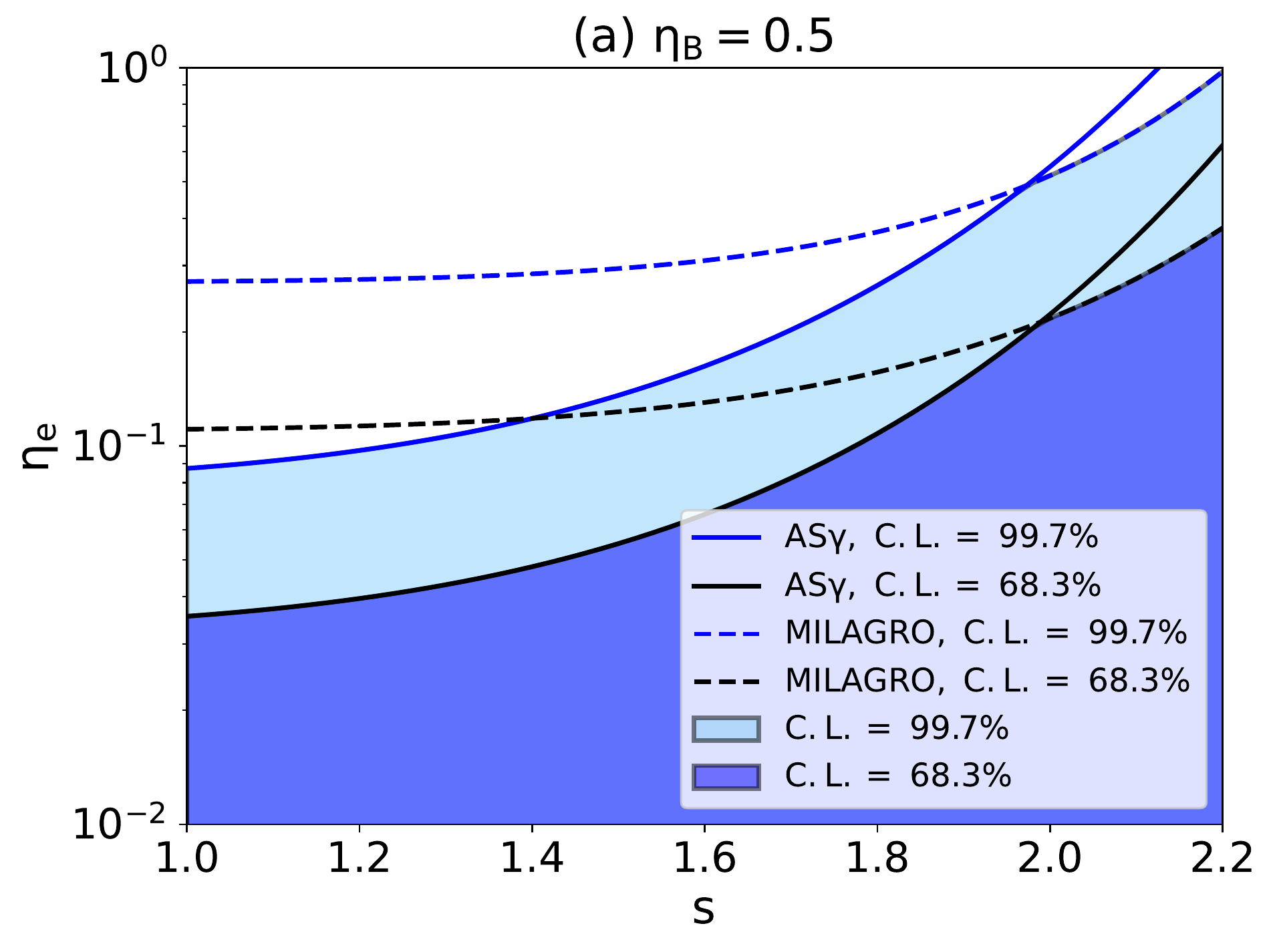}
\includegraphics[width=0.45\textwidth]{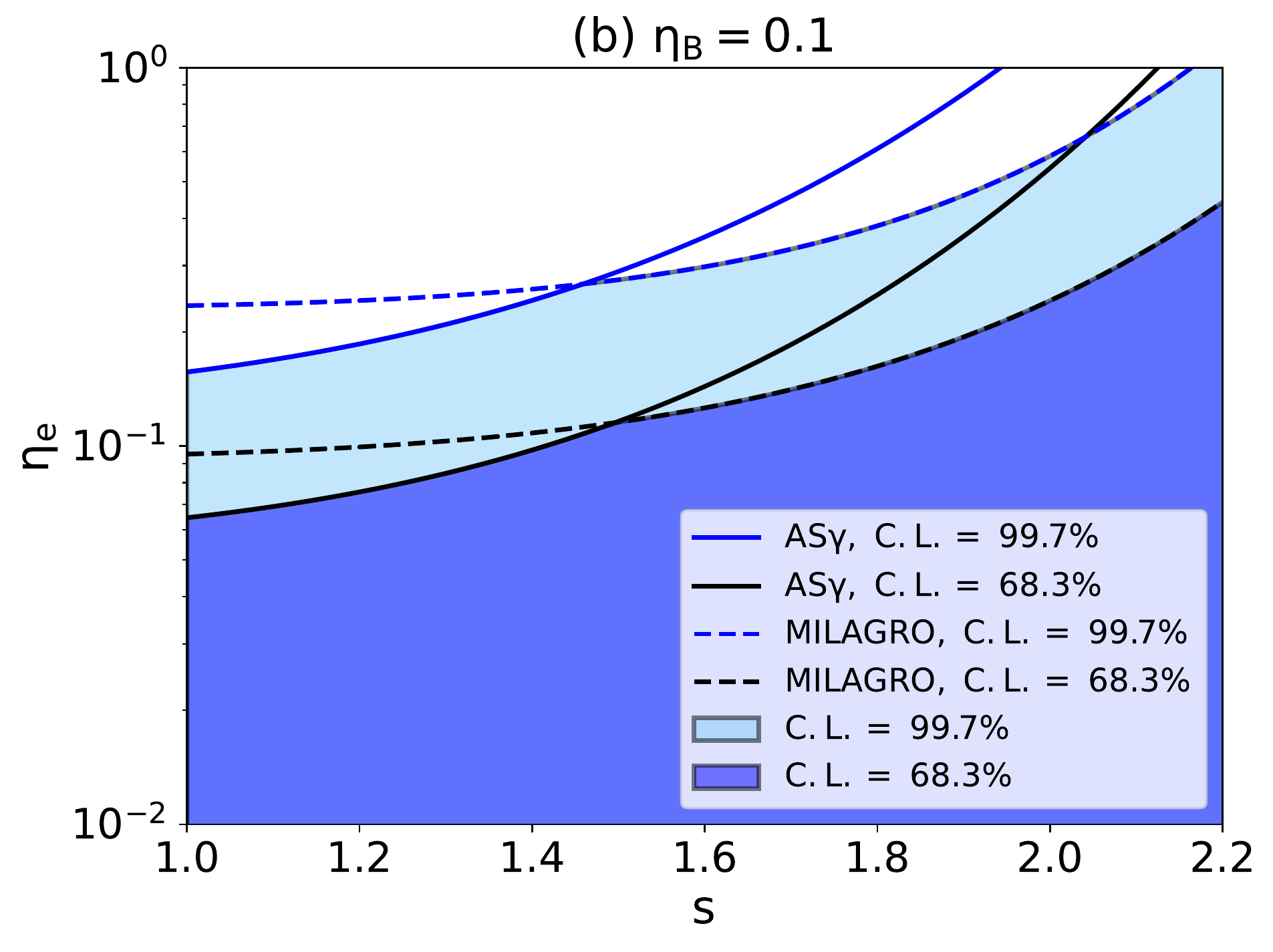}
\caption{Two-dimensional constraints between $s$ and $\eta_{\rm e}$, with $E_{\rm max}$ computed from Eq.~(\ref{eq:etaB}) with (a) $\eta_{B}=0.5$ (b)$\eta_{B}=0.1$. The meaning of graphical elements is the same as in Fig.~\ref{fig:Emax_with}.}
\label{fig:etaB}
\end{figure*}

In Fig.~\ref{fig:etaB}, we present the two-dimensional constraints between $s$ and $\eta_{\rm e}$, with $E_{\rm max}$ computed from Eq.~(\ref{eq:etaB}). In the left panel, we set $\eta_{B}=0.5$ as an optimistic case, while we set $\eta_{B}=0.1$ in the right panel as a conservative case. As expected, we see that the constraints obtained from MILAGRO, i.e. dashed lines, have no significant changes compared with those shown in Fig.~\ref{fig:Emax_with}(b), because of the yielded gamma-ray flux at several TeV does not depend on $E_{\rm max}$ much. On the contrary, the constraints obtained from AS$\gamma$ (solid lines) vary remarkably with changing $\eta_B$. In general, the case of $\eta_B=0.1$ yields a more or less comparable constraint on $\eta_e$ and $s$ to that with a fixed $E_{\rm max}=300\,$TeV. However, it may be worth noting that employing Eq.~\ref{eq:etaB} will increase the contribution from more energetic pulsars while decrease the contribution from less energetic pulsars at the energy band of AS$\gamma$. For $\eta_B=0.1$, only those pulsars with $L_s\gtrsim 2\times 10^{34}\,$erg/s may generate electrons above 100\,TeV. As a result, the resulting gamma-ray flux above 100\,TeV are dominated by a few pulsar halos with the most energetic pulsars. The constraint on the parameters would be less robust in this case since it would be easily influenced by the properties of some most energetic pulsars in the sample.

\subsection{Broken Power-Law Spectrum}

In the previous calculation, we adopt a single power-law injection spectrum with a high-energy exponential cutoff. For relatively soft injection spectra, i.e. $s>2$, the obtained constraints on parameters are quite relaxed. This is because in the cases of $s>$2, the resulting IC flux is mainly concentrated at the GeV band. Some previous literature \cite[e.g.][]{2019Johannesson, 2022Martin} considered a broken power-law form for the injection electron spectrum with the break energy $E_{\rm b}$ usually at $0.1-1\,$TeV. Below the break energy, the spectrum is very hard, which is roughly equivalent to setting the minimum energy $E_{0}$ to $E_b$. As a result, introducing a spectral break in the case of $s>2$ will increase the energies distributed at the high-energy end provided the same total injection luminosity. We then adopt a broken power-law injection spectrum, i.e.
\begin{eqnarray}
Q(E_{\rm e}, t)=N_{0}(t) \times e^{-E_{\rm e} / E_{\rm max}} \times\left\{\begin{array}{ll}
\left(E_{\rm e} / E_{\rm b}\right)^{-s_{1}} & \text { for } E_{\rm e} \leq E_{\rm b} \\
\left(E_{\rm e} / E_{\rm b}\right)^{-s_{2}} & \text { for } E_{\rm e}>E_{\rm b}
\end{array}\right.
\label{eq:broken}
\end{eqnarray}
with $s_2\geq 2$, and repeat the calculation mentioned above.

\begin{figure*}[htbp]
%\hspace{0.9cm} 
\centering
\includegraphics[width=0.45\textwidth]{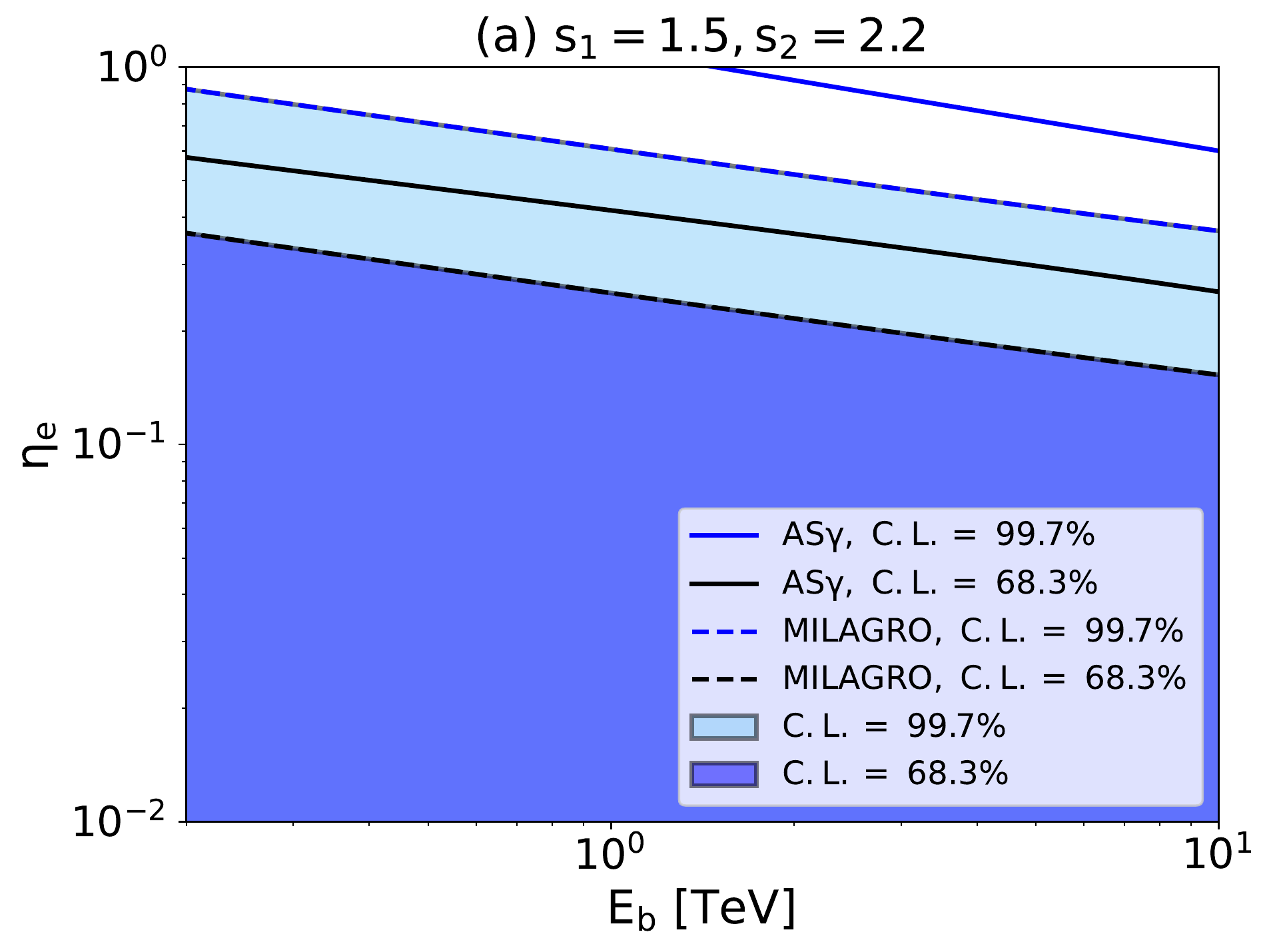}
\includegraphics[width=0.45\textwidth]{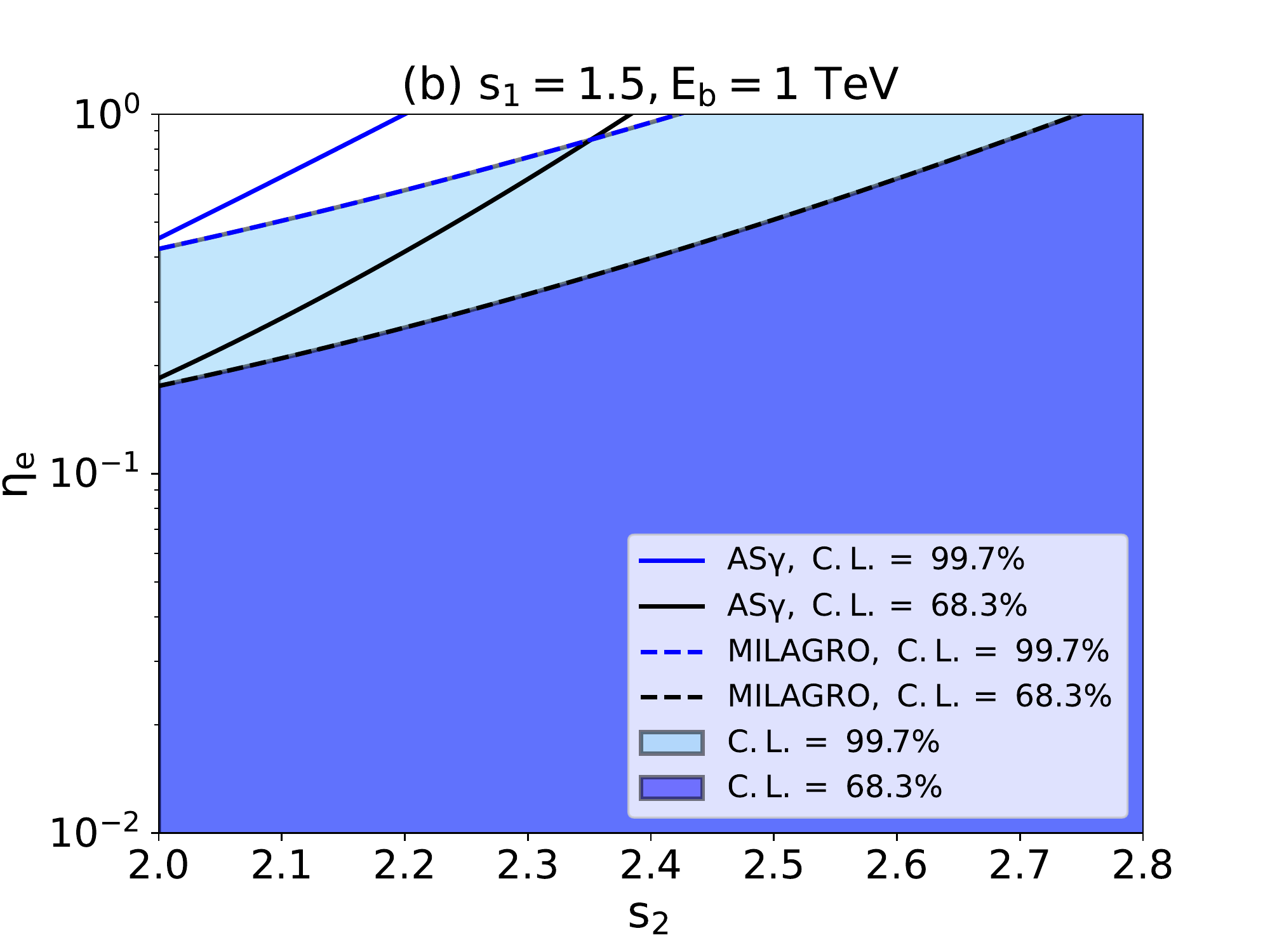}
\caption{(a) two-dimensional constraints between $E_{\rm b}$ and $\eta_{\rm e}$, with $s_1$ fixed to 1.5 and $s_2$ fixed to 2.2 (b) two-dimensional constraints between $s_2$ and $\eta_{\rm e}$, with $s_1$ fixed to 1.5 and $E_{\rm b}$ fixed to $\rm 1~TeV$. The meaning of graphical elements is the same as in Fig.~\ref{fig:Emax_with}.}
\label{fig:soft}
\end{figure*}

In Fig.~\ref{fig:soft}, we present the result in this case. The low energy spectral index is fixed to $s_1=1.5$ and $E_{\rm max}$ is computed from Eq.~(\ref{eq:etaB}) with $\eta_B=0.5$. In the left panel, we fix the high energy spectral index $s_2=2.2$ and explore the effects of $E_{\rm b}$, while we fix $E_{\rm b}=1 \rm TeV$ and vary $s_2$ in the right panel. We see that the MILAGRO data is more constraining than AS$\gamma$ data in the considered parameter space. The constraint is stricter than that in the single power-law case. In general, we can rule out a very large value of $\eta_e$($\sim 1$) unless the injection spectrum is very soft at high energies, i.e., $s_2>2.4$ and $E_b<0.2\,$TeV.

%We see that $\eta_e$ has a tendency to decrease as $E_{\rm b}$ increases. This is because the increase of $E_{\rm b}$ leads to the decrease of photon fluxes at the low energy, while the high energy photon flux is fixed by the observation of MILAGRO and AS$\rm \gamma$. Thus, the total gamma-ray flux is reduced and so does the energy efficiency $\eta_{\rm e}$. 

\subsection{Comparison with Observed Pulsar Halos}
 The results obtained in previous sections represent a general constraint on the particle injection from the pulsar halo population. It may be worth comparing the constraint with the parameters obtained by fitting individual pulsar halos, namely, the halos of the Geminga pulsar, the Monogem pulsar and PSR~J0622+3749. For the former two halos, the measured spectra have large uncertainties and hence the particle injection parameters are not well constrained. Taking the Geminga pulsar as an example, the observed spectrum of its TeV gamma-ray halo can be interpreted with various combinations of $s$ and $\eta_e$. According to previous literature, the injection spectrum could be as hard as $s=1.5-1.6$ with $\eta_e=0.03-0.05$ \citep{Liu19prl, Recchia21}, or as soft as $s=2.3-2.4$ with $\eta_e=0.4-0.6$ \citep{2019Tang, 2022Martin} under the pure diffusion model, which is consistent with the 68.3\% C.L. upper limit as shown in Figs.~\ref{fig:Emax_with} and \ref{fig:etaB}. The fitting to the Monogem's halo results in similar parameters. On the other hand, for the halo of PSR~J0622+3749, the measurements of LHAASO and Fermi-LAT suggest a hard injection spectrum of $s=1.5$ with a high pair conversion efficiency $\eta_e=0.4$ \citep{2021LHAASO}, which exceeds the constraint from the UL of 99.7\% C.L. of the DGE. However, we note that the distance of PSR~J0622+3749 is of large uncertainty. The obtained high pair conversion efficiency is based on a ``pseudo distance'' of $d=1.6$\,kpc derived from the empirical relation between the gamma-ray luminosity and the spin-down luminosity for gamma-ray pulsars \citep{SazParkinson2010}. Provided a smaller distance of the pulsar, the requirement on the pair conversion efficiency would be less stringent since $\eta_e\propto d^2$.

On the other hand, under the model suggested by \citet{Recchia21}, where the standard ISM diffusion coefficient is employed with considering ballistic propagation of injected pairs at small radii, the required pair conversion efficiencies for the halo of Geminga exceeds unity while for the halos of Monogem and J0622+3749 are around unity, even with a hard injection spectrum $s=1.5$. With the parameters, the expected diffuse gamma-ray flux from the entire pulsar halo population would largely overshoot the measurements by MILAGRO and AS$\gamma$, which thus disfavor the model. Otherwise, one might instead argue that only a small fraction of middle-aged pulsars can form pulsar halos around them under this model, but this would raise new questions to be elucidated, such as the special conditions of forming gamma-ray halos for these pulsars.

\section{\label{sec:Sec4}Discussion}

\subsection{Possible improvements to get tighter constraints }\label{sec:sec4a}
The obtained parameter constraints are upper limits rather than exact values. To avoid introducing extra uncertainties, we have simplified some conditions intentionally and obtained conservative upper limits. In other words, one might obtain tighter constraints after carefully considering these effects.

The instrumental detection threshold leads to a missing population of pulsars. Most pulsars are detected in the radio band, while the spin-down luminosity of pulsars do not clearly correlate with the radio luminosity. Therefore, it is possible that some pulsars of relatively high spin-down luminosity are not observable. Taking into account halos of these miss pulsars would lead to a higher expected diffuse gamma-ray flux, which can be translated into stronger constraints on injection parameters. 

\citet{2004Yusifov} derived the correction factor of the intrinsic the pulsar spatial distribution in the Galaxy based on 1400 ATNF pulsar samples at the time. 
Applying for their correction factor in our calculation (see Appendix~\ref{app:c} for details) leads to almost ten-fold higher diffuse gamma-ray flux, and, as a consequence, the obtained upper limit of $\eta_{\rm e}$ would be decreased by about an order of magnitude for any given spectral index $s$ or $E_{\rm max}$, as illustrated in Fig.~\ref{fig:append}. However, the instrumental detectivity of pulsars at nowadays must be advanced with respect to that at the time of \citet{2004Yusifov}, because of improvement of sensitivity of new radio telescopes and  increase of exposure time for recent 20 years. As a result, the amount of sub-threshold pulsars is very likely overestimated by employing the correction factor obtained almost twenty years ago to the pulsar sample at the present time (i.e., some of sub-threshold pulsars at that time would have already been discovered today). Hence, the obtained upper limits for $\eta_e$ in Fig.~\ref{fig:append} would be too radical to be realistic after employing their correction factor. A dedicated study is needed to obtain an accurate correction factor based on the present pulsar sample, which is, however, beyond the scope of this work. In order to make our result and conclusion on the safe side, we choose not to count in the contribution of those sub-threshold pulsars in the present study, noting that our conservative result can provide nontrivial constraints on models of pulsar halos as shown in the previous section. 

\subsection{Uncertainties of the Method}\label{sec:sec4b}
%\textbf{As we discussed in Sec.~\ref{sec:Sec2.1}, characteristic ages of pulsars could deviate from their true ages, which depends on its initial rotation parameters and evolution model. So the pulsar samples passed our criteria might include some young pulsars which have not generated halos. We reassessed the ages of all the samples according to their rotation periods and SNR associations. Some earlier studies proposed that pulsars might be born with similar periods ($\rm \sim  20~ms$; e.g. \citep{1985Lyne}). Thus, pulsars in fast rotation state (especially for those with period $\rm P<100~ms$) might be much younger than its characteristic age. Apart from PSR J1952+3252, there is another pulsar in fast rotation with $\rm P<100~ms$ and high spin-down luminosity ($\rm L_s=9.54\times 10^{35} ~ erg~s^{-1}$), as shown in the left side of Fig.~\ref{fig:append} near the site of J1952+3252. We considered the possibility that this pulsar does not host a halo either, and this will not make much difference to our results but about a tenth of the difference to our parameter constraints. We also retrieved the SNR associations of all the pulsar samples and find three pulsars have potential SNR associations in our remaining samples. All the three pulsars are in relatively low spin-down luminosity and slow rotation state, as shown with star symbols in Fig.~\ref{fig:append}, and will not have much influence to our results.}

As mentioned at the beginning of Section~\ref{sec:Sec3}, we employed the factorized model proposed by \citet{2018Lipari} to estimate the DGE from interactions between CR hadrons and the ISM, which yields a conservative estimate of the DGE flux in the ROI ($l=25^\circ-100^\circ$, $|b|\leq 5^\circ$) compared to the other model proposed in \citet{2018Lipari}. The other one, coined as the "non-factorized" model, takes into account the hardening of CR spectra toward the inner Galactic region as inferred from the DGE measurement by Fermi-LAT \citep{Fermi2016, Yang2016}. The non-factorized model predicts a higher CR-induced DGE flux in TeV -- PeV than the factorized model by a factor of $\lesssim 2$ in the ROI. If we employ the non-factorized model instead, the room left for the contribution of pulsar halos will be smaller, and, as a consequence, tighter constraints of injection parameters could be obtained. It may be worth noting that the CR-induced DGE flux predicted by the non-factorized model already exceeds the 68.3\% UL of the flux in energy bin of $158 - 398$\,TeV measured by AS$\gamma$. If we ignore this energy bin, the obtained upper limits of $\eta_e$ based on data in other energy bins are roughly a factor of 4 lower than those obtained with the factorized model given the same value of $s$. %We compare the factorized model with the uniform CR scenario, which assumes the spectra and intensity of the CR are identical in the Galaxy, and find the differences of the two on our results are less than two times. }

Compared to previous studies which simulated a sample of pulsar halos based on the supernova rate, our method relies on the observed pulsars and the beam fraction of pulsars to estimate the contribution of a complete sample of pulsar halos in our Galaxy. Our calculation is based on the relation between the characteristic age and the beam fraction suggested by Ref.~\cite{1998Tauris}, which was obtained through the distribution of inclination of the magnetic axes of pulsars. The authors also suggested a relation between the rotation period and the beam fraction, which reads $f_{\rm beam}=0.09[\log(P/10{\rm s})]^2+0.03$. Using the latter relation instead would lead to a slight difference (within a factor of 1.5) in the obtained upper limits of $\eta_e$ as shown in Fig.~\ref{fig:Emax_with}-\ref{fig:soft}, which will not influence our conclusion. On the other hand, we note that both relations lead to a mean beaming faction of $\lesssim 0.2$ for the employed middle-aged pulsar sample, implying that the most majority of middle-aged pulsars are invisible to us. Therefore, it might be an issue that whether the properties of observed pulsars can be a good proxy for the entire pulsar population, and thus might introduce corresponding uncertainty to our results.

%Another uncertainty may come from the particle propagation model. Although we argued at the beginning that electrons will deposit most of their energies in the Galactic disk even with the standard ISM diffusion coefficient, the size of the halo would be different for different diffusion coefficients. If the slow diffusion zone truly exists around the pulsar, the size of the pulsar halo would be limited. In this case, exclusion of the contribution of pulsars located within $0.5^{\circ}$ of known TeV sources may be appropriate. On the other hand, the slow diffusion zone does not necessarily exist under some models (e.g., when considering ballistic propagation at small radius\citep{Recchia21} or considering anisotropic diffusion of particles\citep{Liu19prl}). Particles can then propagate to rather far distances, and hence a considerable fraction of the emission may come from radii larger than $0.5^{\circ}$ from the pulsar, especially for nearby pulsars. Apparently, considering the emission from large radius would enhance the expected DGE flux and could make the constraint stronger. A more sophisticated modelling needs to consider the three-dimensional distribution of electrons injected from all the pulsars in the sample. We leave such a study in the future work, noting that the constraints obtained in this work are based on many conservative assumptions and treatments.

Another uncertainty may come from the particle transport model. In the DGE paper of the AS$\rm \gamma$ collaboration, they masked the region within 0.5 degree of all the known TeV sources in order to get rid of the emission of sources\citep{2021ASgamma}. To compare with the measurement of AS$\rm \gamma$, we remove a pulsar from our sample if the pulsar is located within 0.5 degrees of any known TeV source. However, if the diffusion coefficient in the pulsar halo is not suppressed as suggested by \citet{Recchia21}, the angular size of the pulsar halo would be quite large. For example, the cooling timescale of 100\,TeV electrons in 3\,$\rm \mu G$ magnetic field and CMB is about 10\,kyr. The diffusion coefficient would be $\sim \rm 10^{30}\, cm^{2}/s$ for 100\,TeV electron in ISM without suppression. It leads to a diffusion distance (i.e., halo size) of $\sim 100\,$pc, or an angular size of $\rm 2^\circ \, (d/3kpc)^{-1} \, $, where $d$ is the distance of the pulsar. Since the halo size could be larger than $0.5^\circ$, the halos of the pulsars in the masked region could still contribute to the DGE (see details in Appendix~\ref{app:a}), especially for those nearby pulsars. In this sense, we may have underestimated the DGE flux by simply ignoring those pulsars in the masked region. 
To accurately estimate the flux of these pulsar halos contributed to DGE, a more sophisticated modelling is needed, e.g., considering the position of each pulsar in the Galaxy and the three-dimensional distribution of injected pairs. We leave such a study in the future work, noting that the current treatments lead to conservative constraints.

\subsection{Constraints from the CR positron flux}\label{sec:sec4c}
In addition to DGE, cosmic-ray positron flux might be an alternative observable quantity to constrain injection parameters of pulsar halos, especially in the case of soft injection spectrum. Some previous studies have investigated the expectation of positron flux from middle-aged pulsars \citep{Fang19, Manconi20, 2022Martin}. However, we note that the expected positron flux at Earth is highly dependent on the transport mechanism of particles in pulsar halos. Besides, the positron excess appears at sub-TeV energies, at which positrons injected at early time of a middle-aged pulsar have not been cooled in typical interstellar magnetic field at present time and could make contribution to the position flux at Earth. On the other hand, the amount of these early injected positrons highly depend on the initial rotation period of the pulsar $P_0$ (which determines the initial spindown luminosity) and the electromagnetic environment of the PWN, either of which are unknown. Thus the predicted positron flux at sub-TeV energies could have large uncertainties, preventing us from drawing a concrete conclusion. This is similar to the reason why we do not use the DGE measured by ARGO to constrain models, as illustrated in the left panel of Fig.~\ref{fig:spec}.

\section{\label{sec:Sec5}Summary and Conclusion}
Detection of TeV pulsar halos around middle-aged pulsars suggests escaping of energetic electrons from PWNe of these middle-aged pulsars. It backs up the idea that middle-aged pulsars may be important sources of cosmic-ray leptons, especially for positrons. Therefore, the gamma-ray flux of pulsar halos can reveal the properties of pair injection from the PWNe, as well as the transport mechanism of particles in ISM. If each middle-aged pulsar can form a TeV halo, most of them are not sufficiently bright to be resolved by current instruments and hence their emission may be counted as diffuse gamma-ray emission in the Galactic plane. 

In this work, we tried to acquire constraints on the pair injection parameters of pulsar halos by the DGE flux measured by AS$\rm \gamma$ and MILAGRO. We firstly screened out a sample of middle-aged pulsars within the ROI of AS$\rm \gamma$ and MILAGRO, excluding those pulsars located within $0.5^\circ$ of known TeV sources. We then calculated the expected diffuse gamma-ray emission generated by escaping electrons from these pulsars under a set of injection parameters. We also estimated diffuse gamma-ray flux from the pionic emission of cosmic-ray hadrons, based on the ``factorized'' model developed by Ref.~\cite{2018Lipari}, which likely yields a conservative estimation of the pionic component. We then considered the measured DGE as upper limits of gamma-ray fluxes generated by the sum of cosmic-ray hadrons and unresolved pulsar halos. By comparing the measurements with theoretical predictions, we obtained the constraint on injection
parameters of the pulsar halos such as the pair conversion efficiency $\eta_e$, the injection spectral index $s$ and the maximum cutoff energy $E_{\rm max}$ of injected electrons. More specifically, under a moderate assumption of $E_{\rm max}$, the 68.3\% upper limit of DGE data gives $\eta_{e}\lesssim 0.1$ for $s<1.8$. For a softer spectrum
(i.e., a larger $s$), the MILAGRO data at several TeV energies becomes more constraining, but the overall constraints become less stringent. For $s=2.2$, the upper limit of $\eta_e$ increases to 0.4. In the case that a
hardening appears in the injection spectrum below sub-TeV to a few TeV, a stricter constraint on $\eta_e$ can be obtained given the same spectrum at the high-energy end. 

Based on our results, we may draw the following conclusions:
\begin{itemize}
    \item The expected diffuse gamma-ray emission from unresolved pulsar halos may saturate the measured DGE flux above several TeV with reasonable parameters. Therefore, unresolved pulsar halos may likely contribute a non-negligible fraction of the measured DGE, which is consistent with the result of Ref.~\cite{2018Linden}.
    \item The large value of the pair conversion efficiency $\eta_e\sim 1$ can be generally excluded unless the injection spectrum is very soft. Our result disfavors the isotropic, unsuppressed diffusion model\cite{Recchia21} which requires a hard injection spectrum and a large conversion efficiency $\eta_e\sim 1$.
    \item We reiterate that the obtained constraints are generally in the conservative side because of our intentional choice of some steps in the calculation. However, some uncertainties such as the pulsar's beam fraction might nevertheless influence the results. The future measurement of LHAASO on DGE may give preciser constraints.
\end{itemize}

\begin{acknowledgments}
We thank anonymous referees for constructive reports which improve the quality of the paper. This work is supported by NSFC under the grant No.2031105.
\end{acknowledgments}

\bibliographystyle{apsrev}
\bibliography{apssamp}% Produces the bibliography via BibTeX.

\providecommand{\noopsort}[1]{}\providecommand{\singleletter}[1]{#1}%
\begin{thebibliography}{49}
\expandafter\ifx\csname natexlab\endcsname\relax\def\natexlab#1{#1}\fi
\expandafter\ifx\csname bibnamefont\endcsname\relax
  \def\bibnamefont#1{#1}\fi
\expandafter\ifx\csname bibfnamefont\endcsname\relax
  \def\bibfnamefont#1{#1}\fi
\expandafter\ifx\csname citenamefont\endcsname\relax
  \def\citenamefont#1{#1}\fi
\expandafter\ifx\csname url\endcsname\relax
  \def\url#1{\texttt{#1}}\fi
\expandafter\ifx\csname urlprefix\endcsname\relax\def\urlprefix{URL }\fi
\providecommand{\bibinfo}[2]{#2}
\providecommand{\eprint}[2][]{\url{#2}}

\bibitem[{\citenamefont{{Ackermann} et~al.}(2012)\citenamefont{{Ackermann},
  {Ajello}, {Atwood}, {Baldini}, {Ballet}, {Barbiellini}, {Bastieri},
  {Bechtol}, {Bellazzini}, {Berenji} et~al.}}]{2012Fermi}
\bibinfo{author}{\bibfnamefont{M.}~\bibnamefont{{Ackermann}}},
  \bibinfo{author}{\bibfnamefont{M.}~\bibnamefont{{Ajello}}},
  \bibinfo{author}{\bibfnamefont{W.~B.} \bibnamefont{{Atwood}}},
  \bibinfo{author}{\bibfnamefont{L.}~\bibnamefont{{Baldini}}},
  \bibinfo{author}{\bibfnamefont{J.}~\bibnamefont{{Ballet}}},
  \bibinfo{author}{\bibfnamefont{G.}~\bibnamefont{{Barbiellini}}},
  \bibinfo{author}{\bibfnamefont{D.}~\bibnamefont{{Bastieri}}},
  \bibinfo{author}{\bibfnamefont{K.}~\bibnamefont{{Bechtol}}},
  \bibinfo{author}{\bibfnamefont{R.}~\bibnamefont{{Bellazzini}}},
  \bibinfo{author}{\bibfnamefont{B.}~\bibnamefont{{Berenji}}},
  \bibnamefont{et~al.}, \bibinfo{journal}{\apj} \textbf{\bibinfo{volume}{750}},
  \bibinfo{eid}{3} (\bibinfo{year}{2012}), \eprint{1202.4039}.

\bibitem[{\citenamefont{{Neronov} and {Semikoz}}(2020)}]{2020Neronov}
\bibinfo{author}{\bibfnamefont{A.}~\bibnamefont{{Neronov}}} \bibnamefont{and}
  \bibinfo{author}{\bibfnamefont{D.}~\bibnamefont{{Semikoz}}},
  \bibinfo{journal}{\aap} \textbf{\bibinfo{volume}{633}}, \bibinfo{eid}{A94}
  (\bibinfo{year}{2020}), \eprint{1907.06061}.

\bibitem[{\citenamefont{{Atkins} et~al.}(2005)\citenamefont{{Atkins}, {Benbow},
  {Berley}, {Blaufuss}, {Coyne}, {De Young}, {Dingus}, {Dorfan}, {Ellsworth},
  {Fleysher} et~al.}}]{2005MILAGRO}
\bibinfo{author}{\bibfnamefont{R.}~\bibnamefont{{Atkins}}},
  \bibinfo{author}{\bibfnamefont{W.}~\bibnamefont{{Benbow}}},
  \bibinfo{author}{\bibfnamefont{D.}~\bibnamefont{{Berley}}},
  \bibinfo{author}{\bibfnamefont{E.}~\bibnamefont{{Blaufuss}}},
  \bibinfo{author}{\bibfnamefont{D.~G.} \bibnamefont{{Coyne}}},
  \bibinfo{author}{\bibfnamefont{T.}~\bibnamefont{{De Young}}},
  \bibinfo{author}{\bibfnamefont{B.~L.} \bibnamefont{{Dingus}}},
  \bibinfo{author}{\bibfnamefont{D.~E.} \bibnamefont{{Dorfan}}},
  \bibinfo{author}{\bibfnamefont{R.~W.} \bibnamefont{{Ellsworth}}},
  \bibinfo{author}{\bibfnamefont{L.}~\bibnamefont{{Fleysher}}},
  \bibnamefont{et~al.}, \bibinfo{journal}{\prl} \textbf{\bibinfo{volume}{95}},
  \bibinfo{eid}{251103} (\bibinfo{year}{2005}), \eprint{astro-ph/0502303}.

\bibitem[{\citenamefont{{Bartoli} et~al.}(2015)\citenamefont{{Bartoli},
  {Bernardini}, {Bi}, {Branchini}, {Budano}, {Camarri}, {Cao}, {Cardarelli},
  {Catalanotti}, {Chen} et~al.}}]{2015ARGO}
\bibinfo{author}{\bibfnamefont{B.}~\bibnamefont{{Bartoli}}},
  \bibinfo{author}{\bibfnamefont{P.}~\bibnamefont{{Bernardini}}},
  \bibinfo{author}{\bibfnamefont{X.~J.} \bibnamefont{{Bi}}},
  \bibinfo{author}{\bibfnamefont{P.}~\bibnamefont{{Branchini}}},
  \bibinfo{author}{\bibfnamefont{A.}~\bibnamefont{{Budano}}},
  \bibinfo{author}{\bibfnamefont{P.}~\bibnamefont{{Camarri}}},
  \bibinfo{author}{\bibfnamefont{Z.}~\bibnamefont{{Cao}}},
  \bibinfo{author}{\bibfnamefont{R.}~\bibnamefont{{Cardarelli}}},
  \bibinfo{author}{\bibfnamefont{S.}~\bibnamefont{{Catalanotti}}},
  \bibinfo{author}{\bibfnamefont{S.~Z.} \bibnamefont{{Chen}}},
  \bibnamefont{et~al.}, \bibinfo{journal}{\apj} \textbf{\bibinfo{volume}{806}},
  \bibinfo{eid}{20} (\bibinfo{year}{2015}), \eprint{1507.06758}.

\bibitem[{\citenamefont{{Amenomori} et~al.}(2021)\citenamefont{{Amenomori},
  {Bao}, {Bi}, {Chen}, {Chen}, {Chen}, {Chen}, {Chen}, {Cirennima},
  {Danzengluobu} et~al.}}]{2021ASgamma}
\bibinfo{author}{\bibfnamefont{M.}~\bibnamefont{{Amenomori}}},
  \bibinfo{author}{\bibfnamefont{Y.~W.} \bibnamefont{{Bao}}},
  \bibinfo{author}{\bibfnamefont{X.~J.} \bibnamefont{{Bi}}},
  \bibinfo{author}{\bibfnamefont{D.}~\bibnamefont{{Chen}}},
  \bibinfo{author}{\bibfnamefont{T.~L.} \bibnamefont{{Chen}}},
  \bibinfo{author}{\bibfnamefont{W.~Y.} \bibnamefont{{Chen}}},
  \bibinfo{author}{\bibfnamefont{X.}~\bibnamefont{{Chen}}},
  \bibinfo{author}{\bibfnamefont{Y.}~\bibnamefont{{Chen}}},
  \bibinfo{author}{\bibfnamefont{S.~W.} \bibnamefont{{Cirennima}},
  \bibfnamefont{Cui}}, \bibinfo{author}{\bibfnamefont{L.~K.}
  \bibnamefont{{Danzengluobu}}, \bibfnamefont{Ding}}, \bibnamefont{et~al.},
  \bibinfo{journal}{\prl} \textbf{\bibinfo{volume}{126}}, \bibinfo{eid}{141101}
  (\bibinfo{year}{2021}), \eprint{2104.05181}.

\bibitem[{\citenamefont{{Linden} and {Buckman}}(2018)}]{2018Linden}
\bibinfo{author}{\bibfnamefont{T.}~\bibnamefont{{Linden}}} \bibnamefont{and}
  \bibinfo{author}{\bibfnamefont{B.~J.} \bibnamefont{{Buckman}}},
  \bibinfo{journal}{\prl} \textbf{\bibinfo{volume}{120}}, \bibinfo{eid}{121101}
  (\bibinfo{year}{2018}), \eprint{1707.01905}.

\bibitem[{\citenamefont{{Liu} and {Wang}}(2021)}]{21Liu}
\bibinfo{author}{\bibfnamefont{R.-Y.} \bibnamefont{{Liu}}} \bibnamefont{and}
  \bibinfo{author}{\bibfnamefont{X.-Y.} \bibnamefont{{Wang}}},
  \bibinfo{journal}{\apjl} \textbf{\bibinfo{volume}{914}}, \bibinfo{eid}{L7}
  (\bibinfo{year}{2021}), \eprint{2104.05609}.

\bibitem[{\citenamefont{{Fang} et~al.}(2022)\citenamefont{{Fang}, {Gallagher},
  and {Halzen}}}]{2022Fang}
\bibinfo{author}{\bibfnamefont{K.}~\bibnamefont{{Fang}}},
  \bibinfo{author}{\bibfnamefont{J.~S.} \bibnamefont{{Gallagher}}},
  \bibnamefont{and} \bibinfo{author}{\bibfnamefont{F.}~\bibnamefont{{Halzen}}},
  \bibinfo{journal}{\apj} \textbf{\bibinfo{volume}{933}}, \bibinfo{eid}{190}
  (\bibinfo{year}{2022}), \eprint{2205.03740}.

\bibitem[{\citenamefont{{Vecchiotti}
  et~al.}(2022{\natexlab{a}})\citenamefont{{Vecchiotti}, {Pagliaroli}, and
  {Villante}}}]{Vecchiotti22a}
\bibinfo{author}{\bibfnamefont{V.}~\bibnamefont{{Vecchiotti}}},
  \bibinfo{author}{\bibfnamefont{G.}~\bibnamefont{{Pagliaroli}}},
  \bibnamefont{and} \bibinfo{author}{\bibfnamefont{F.~L.}
  \bibnamefont{{Villante}}}, \bibinfo{journal}{Communications Physics}
  \textbf{\bibinfo{volume}{5}}, \bibinfo{eid}{161}
  (\bibinfo{year}{2022}{\natexlab{a}}), \eprint{2107.03236}.

\bibitem[{\citenamefont{{Vecchiotti}
  et~al.}(2022{\natexlab{b}})\citenamefont{{Vecchiotti}, {Zuccarini},
  {Villante}, and {Pagliaroli}}}]{Vecchiotti22b}
\bibinfo{author}{\bibfnamefont{V.}~\bibnamefont{{Vecchiotti}}},
  \bibinfo{author}{\bibfnamefont{F.}~\bibnamefont{{Zuccarini}}},
  \bibinfo{author}{\bibfnamefont{F.~L.} \bibnamefont{{Villante}}},
  \bibnamefont{and}
  \bibinfo{author}{\bibfnamefont{G.}~\bibnamefont{{Pagliaroli}}},
  \bibinfo{journal}{\apj} \textbf{\bibinfo{volume}{928}}, \bibinfo{eid}{19}
  (\bibinfo{year}{2022}{\natexlab{b}}), \eprint{2107.14584}.

\bibitem[{\citenamefont{{L{\'o}pez-Coto}
  et~al.}(2022)\citenamefont{{L{\'o}pez-Coto}, {de O{\~n}a Wilhelmi},
  {Aharonian}, {Amato}, and {Hinton}}}]{2022Lopez-Coto}
\bibinfo{author}{\bibfnamefont{R.}~\bibnamefont{{L{\'o}pez-Coto}}},
  \bibinfo{author}{\bibfnamefont{E.}~\bibnamefont{{de O{\~n}a Wilhelmi}}},
  \bibinfo{author}{\bibfnamefont{F.}~\bibnamefont{{Aharonian}}},
  \bibinfo{author}{\bibfnamefont{E.}~\bibnamefont{{Amato}}}, \bibnamefont{and}
  \bibinfo{author}{\bibfnamefont{J.}~\bibnamefont{{Hinton}}},
  \bibinfo{journal}{Nature Astronomy} \textbf{\bibinfo{volume}{6}},
  \bibinfo{pages}{199} (\bibinfo{year}{2022}), \eprint{2202.06899}.

\bibitem[{\citenamefont{{Liu}}(2022)}]{22Liu}
\bibinfo{author}{\bibfnamefont{R.-Y.} \bibnamefont{{Liu}}},
  \bibinfo{journal}{arXiv e-prints} \bibinfo{eid}{arXiv:2207.04011}
  (\bibinfo{year}{2022}), \eprint{2207.04011}.

\bibitem[{\citenamefont{{Abeysekara} et~al.}(2017)\citenamefont{{Abeysekara},
  {Albert}, {Alfaro}, {Alvarez}, {{\'A}lvarez}, {Arceo},
  {Arteaga-Vel{\'a}zquez}, {Avila Rojas}, {Ayala Solares}, {Barber}
  et~al.}}]{2017Abeysekara}
\bibinfo{author}{\bibfnamefont{A.~U.} \bibnamefont{{Abeysekara}}},
  \bibinfo{author}{\bibfnamefont{A.}~\bibnamefont{{Albert}}},
  \bibinfo{author}{\bibfnamefont{R.}~\bibnamefont{{Alfaro}}},
  \bibinfo{author}{\bibfnamefont{C.}~\bibnamefont{{Alvarez}}},
  \bibinfo{author}{\bibfnamefont{J.~D.} \bibnamefont{{{\'A}lvarez}}},
  \bibinfo{author}{\bibfnamefont{R.}~\bibnamefont{{Arceo}}},
  \bibinfo{author}{\bibfnamefont{J.~C.} \bibnamefont{{Arteaga-Vel{\'a}zquez}}},
  \bibinfo{author}{\bibfnamefont{D.}~\bibnamefont{{Avila Rojas}}},
  \bibinfo{author}{\bibfnamefont{H.~A.} \bibnamefont{{Ayala Solares}}},
  \bibinfo{author}{\bibfnamefont{A.~S.} \bibnamefont{{Barber}}},
  \bibnamefont{et~al.}, \bibinfo{journal}{Science}
  \textbf{\bibinfo{volume}{358}}, \bibinfo{pages}{911} (\bibinfo{year}{2017}),
  \eprint{1711.06223}.

\bibitem[{\citenamefont{{Aharonian} et~al.}(2021)\citenamefont{{Aharonian},
  {An}, {Axikegu}, {Bai}, {Bao}, {Bastieri}, {Bi}, {Bi}, {Cai}, {Cai}
  et~al.}}]{2021LHAASO}
\bibinfo{author}{\bibfnamefont{F.}~\bibnamefont{{Aharonian}}},
  \bibinfo{author}{\bibfnamefont{Q.}~\bibnamefont{{An}}},
  \bibinfo{author}{\bibfnamefont{L.~X.} \bibnamefont{{Axikegu}},
  \bibfnamefont{Bai}}, \bibinfo{author}{\bibfnamefont{Y.~X.}
  \bibnamefont{{Bai}}}, \bibinfo{author}{\bibfnamefont{Y.~W.}
  \bibnamefont{{Bao}}},
  \bibinfo{author}{\bibfnamefont{D.}~\bibnamefont{{Bastieri}}},
  \bibinfo{author}{\bibfnamefont{X.~J.} \bibnamefont{{Bi}}},
  \bibinfo{author}{\bibfnamefont{Y.~J.} \bibnamefont{{Bi}}},
  \bibinfo{author}{\bibfnamefont{H.}~\bibnamefont{{Cai}}},
  \bibinfo{author}{\bibfnamefont{J.~T.} \bibnamefont{{Cai}}},
  \bibnamefont{et~al.}, \bibinfo{journal}{\prl} \textbf{\bibinfo{volume}{126}},
  \bibinfo{eid}{241103} (\bibinfo{year}{2021}), \eprint{2106.09396}.

\bibitem[{\citenamefont{{Liu} et~al.}(2019)\citenamefont{{Liu}, {Yan}, and
  {Zhang}}}]{Liu19prl}
\bibinfo{author}{\bibfnamefont{R.-Y.} \bibnamefont{{Liu}}},
  \bibinfo{author}{\bibfnamefont{H.}~\bibnamefont{{Yan}}}, \bibnamefont{and}
  \bibinfo{author}{\bibfnamefont{H.}~\bibnamefont{{Zhang}}},
  \bibinfo{journal}{\prl} \textbf{\bibinfo{volume}{123}}, \bibinfo{eid}{221103}
  (\bibinfo{year}{2019}), \eprint{1904.11536}.

\bibitem[{\citenamefont{{Recchia} et~al.}(2021)\citenamefont{{Recchia}, {Di
  Mauro}, {Aharonian}, {Orusa}, {Donato}, {Gabici}, and {Manconi}}}]{Recchia21}
\bibinfo{author}{\bibfnamefont{S.}~\bibnamefont{{Recchia}}},
  \bibinfo{author}{\bibfnamefont{M.}~\bibnamefont{{Di Mauro}}},
  \bibinfo{author}{\bibfnamefont{F.~A.} \bibnamefont{{Aharonian}}},
  \bibinfo{author}{\bibfnamefont{L.}~\bibnamefont{{Orusa}}},
  \bibinfo{author}{\bibfnamefont{F.}~\bibnamefont{{Donato}}},
  \bibinfo{author}{\bibfnamefont{S.}~\bibnamefont{{Gabici}}}, \bibnamefont{and}
  \bibinfo{author}{\bibfnamefont{S.}~\bibnamefont{{Manconi}}},
  \bibinfo{journal}{\prd} \textbf{\bibinfo{volume}{104}}, \bibinfo{eid}{123017}
  (\bibinfo{year}{2021}), \eprint{2106.02275}.

\bibitem[{\citenamefont{{Yan} et~al.}(2022)\citenamefont{{Yan}, {Liu}, {Chen},
  and {Wang}}}]{Yan22}
\bibinfo{author}{\bibfnamefont{K.}~\bibnamefont{{Yan}}},
  \bibinfo{author}{\bibfnamefont{R.-Y.} \bibnamefont{{Liu}}},
  \bibinfo{author}{\bibfnamefont{S.~Z.} \bibnamefont{{Chen}}},
  \bibnamefont{and} \bibinfo{author}{\bibfnamefont{X.-Y.}
  \bibnamefont{{Wang}}}, \bibinfo{journal}{\apj}
  \textbf{\bibinfo{volume}{935}}, \bibinfo{eid}{65} (\bibinfo{year}{2022}),
  \eprint{2205.14563}.

\bibitem[{\citenamefont{{De La Torre Luque} et~al.}(2022)\citenamefont{{De La
  Torre Luque}, {Fornieri}, and {Linden}}}]{Luque22}
\bibinfo{author}{\bibfnamefont{P.}~\bibnamefont{{De La Torre Luque}}},
  \bibinfo{author}{\bibfnamefont{O.}~\bibnamefont{{Fornieri}}},
  \bibnamefont{and} \bibinfo{author}{\bibfnamefont{T.}~\bibnamefont{{Linden}}},
  \bibinfo{journal}{arXiv e-prints} \bibinfo{eid}{arXiv:2205.08544}
  (\bibinfo{year}{2022}), \eprint{2205.08544}.

\bibitem[{\citenamefont{{Bao} et~al.}(2022)\citenamefont{{Bao}, {Fang}, {Bi},
  and {Wang}}}]{Bao22}
\bibinfo{author}{\bibfnamefont{L.-Z.} \bibnamefont{{Bao}}},
  \bibinfo{author}{\bibfnamefont{K.}~\bibnamefont{{Fang}}},
  \bibinfo{author}{\bibfnamefont{X.-J.} \bibnamefont{{Bi}}}, \bibnamefont{and}
  \bibinfo{author}{\bibfnamefont{S.-H.} \bibnamefont{{Wang}}},
  \bibinfo{journal}{\apj} \textbf{\bibinfo{volume}{936}}, \bibinfo{eid}{183}
  (\bibinfo{year}{2022}), \eprint{2107.07395}.

\bibitem[{\citenamefont{{Trotta} et~al.}(2011)\citenamefont{{Trotta},
  {J{\'o}hannesson}, {Moskalenko}, {Porter}, {Ruiz de Austri}, and
  {Strong}}}]{11Trotta}
\bibinfo{author}{\bibfnamefont{R.}~\bibnamefont{{Trotta}}},
  \bibinfo{author}{\bibfnamefont{G.}~\bibnamefont{{J{\'o}hannesson}}},
  \bibinfo{author}{\bibfnamefont{I.~V.} \bibnamefont{{Moskalenko}}},
  \bibinfo{author}{\bibfnamefont{T.~A.} \bibnamefont{{Porter}}},
  \bibinfo{author}{\bibfnamefont{R.}~\bibnamefont{{Ruiz de Austri}}},
  \bibnamefont{and} \bibinfo{author}{\bibfnamefont{A.~W.}
  \bibnamefont{{Strong}}}, \bibinfo{journal}{\apj}
  \textbf{\bibinfo{volume}{729}}, \bibinfo{eid}{106} (\bibinfo{year}{2011}),
  \eprint{1011.0037}.

\bibitem[{\citenamefont{{Manchester} et~al.}(2005)\citenamefont{{Manchester},
  {Hobbs}, {Teoh}, and {Hobbs}}}]{2005Manchester}
\bibinfo{author}{\bibfnamefont{R.~N.} \bibnamefont{{Manchester}}},
  \bibinfo{author}{\bibfnamefont{G.~B.} \bibnamefont{{Hobbs}}},
  \bibinfo{author}{\bibfnamefont{A.}~\bibnamefont{{Teoh}}}, \bibnamefont{and}
  \bibinfo{author}{\bibfnamefont{M.}~\bibnamefont{{Hobbs}}},
  \bibinfo{journal}{\aj} \textbf{\bibinfo{volume}{129}}, \bibinfo{pages}{1993}
  (\bibinfo{year}{2005}), \eprint{astro-ph/0412641}.

\bibitem[{\citenamefont{{Wakely} and {Horan}}(2008)}]{2008TeVcat}
\bibinfo{author}{\bibfnamefont{S.~P.} \bibnamefont{{Wakely}}} \bibnamefont{and}
  \bibinfo{author}{\bibfnamefont{D.}~\bibnamefont{{Horan}}}, in
  \emph{\bibinfo{booktitle}{International Cosmic Ray Conference}}
  (\bibinfo{year}{2008}), vol.~\bibinfo{volume}{3} of
  \emph{\bibinfo{series}{International Cosmic Ray Conference}}, pp.
  \bibinfo{pages}{1341--1344}.

\bibitem[{\citenamefont{{Hooper} and {Linden}}(2022)}]{22Hooper}
\bibinfo{author}{\bibfnamefont{D.}~\bibnamefont{{Hooper}}} \bibnamefont{and}
  \bibinfo{author}{\bibfnamefont{T.}~\bibnamefont{{Linden}}},
  \bibinfo{journal}{\prd} \textbf{\bibinfo{volume}{105}}, \bibinfo{eid}{103013}
  (\bibinfo{year}{2022}), \eprint{2104.00014}.

\bibitem[{\citenamefont{{Bednarek} and {Bartosik}}(2003)}]{2003Bednarek}
\bibinfo{author}{\bibfnamefont{W.}~\bibnamefont{{Bednarek}}} \bibnamefont{and}
  \bibinfo{author}{\bibfnamefont{M.}~\bibnamefont{{Bartosik}}},
  \bibinfo{journal}{\aap} \textbf{\bibinfo{volume}{405}}, \bibinfo{pages}{689}
  (\bibinfo{year}{2003}), \eprint{astro-ph/0304049}.

\bibitem[{\citenamefont{{Khangulyan} et~al.}(2014)\citenamefont{{Khangulyan},
  {Aharonian}, and {Kelner}}}]{2014Khangulyan}
\bibinfo{author}{\bibfnamefont{D.}~\bibnamefont{{Khangulyan}}},
  \bibinfo{author}{\bibfnamefont{F.~A.} \bibnamefont{{Aharonian}}},
  \bibnamefont{and} \bibinfo{author}{\bibfnamefont{S.~R.}
  \bibnamefont{{Kelner}}}, \bibinfo{journal}{\apj}
  \textbf{\bibinfo{volume}{783}}, \bibinfo{eid}{100} (\bibinfo{year}{2014}),
  \eprint{1310.7971}.

\bibitem[{\citenamefont{{Fouka} and {Ouichaoui}}(2013)}]{2013Fouka}
\bibinfo{author}{\bibfnamefont{M.}~\bibnamefont{{Fouka}}} \bibnamefont{and}
  \bibinfo{author}{\bibfnamefont{S.}~\bibnamefont{{Ouichaoui}}},
  \bibinfo{journal}{Research in Astronomy and Astrophysics}
  \textbf{\bibinfo{volume}{13}}, \bibinfo{eid}{680-686} (\bibinfo{year}{2013}),
  \eprint{1301.6908}.

\bibitem[{\citenamefont{{Tauris} and {Manchester}}(1998)}]{1998Tauris}
\bibinfo{author}{\bibfnamefont{T.~M.} \bibnamefont{{Tauris}}} \bibnamefont{and}
  \bibinfo{author}{\bibfnamefont{R.~N.} \bibnamefont{{Manchester}}},
  \bibinfo{journal}{\mnras} \textbf{\bibinfo{volume}{298}},
  \bibinfo{pages}{625} (\bibinfo{year}{1998}).

\bibitem[{\citenamefont{{Atoyan} and {Aharonian}}(1996)}]{1996Atoyan}
\bibinfo{author}{\bibfnamefont{A.~M.} \bibnamefont{{Atoyan}}} \bibnamefont{and}
  \bibinfo{author}{\bibfnamefont{F.~A.} \bibnamefont{{Aharonian}}},
  \bibinfo{journal}{\mnras} \textbf{\bibinfo{volume}{278}},
  \bibinfo{pages}{525} (\bibinfo{year}{1996}).

\bibitem[{\citenamefont{{Kothes} et~al.}(2006)\citenamefont{{Kothes}, {Reich},
  and {Uyan{\i}ker}}}]{2006Kothes}
\bibinfo{author}{\bibfnamefont{R.}~\bibnamefont{{Kothes}}},
  \bibinfo{author}{\bibfnamefont{W.}~\bibnamefont{{Reich}}}, \bibnamefont{and}
  \bibinfo{author}{\bibfnamefont{B.}~\bibnamefont{{Uyan{\i}ker}}},
  \bibinfo{journal}{\apj} \textbf{\bibinfo{volume}{638}}, \bibinfo{pages}{225}
  (\bibinfo{year}{2006}).

\bibitem[{\citenamefont{{Gelfand} et~al.}(2009)\citenamefont{{Gelfand},
  {Slane}, and {Zhang}}}]{2009Gelfand}
\bibinfo{author}{\bibfnamefont{J.~D.} \bibnamefont{{Gelfand}}},
  \bibinfo{author}{\bibfnamefont{P.~O.} \bibnamefont{{Slane}}},
  \bibnamefont{and} \bibinfo{author}{\bibfnamefont{W.}~\bibnamefont{{Zhang}}},
  \bibinfo{journal}{\apj} \textbf{\bibinfo{volume}{703}}, \bibinfo{pages}{2051}
  (\bibinfo{year}{2009}), \eprint{0904.4053}.

\bibitem[{\citenamefont{{Hinton} et~al.}(2011)\citenamefont{{Hinton}, {Funk},
  {Parsons}, and {Ohm}}}]{2011Hinton}
\bibinfo{author}{\bibfnamefont{J.~A.} \bibnamefont{{Hinton}}},
  \bibinfo{author}{\bibfnamefont{S.}~\bibnamefont{{Funk}}},
  \bibinfo{author}{\bibfnamefont{R.~D.} \bibnamefont{{Parsons}}},
  \bibnamefont{and} \bibinfo{author}{\bibfnamefont{S.}~\bibnamefont{{Ohm}}},
  \bibinfo{journal}{\apjl} \textbf{\bibinfo{volume}{743}}, \bibinfo{eid}{L7}
  (\bibinfo{year}{2011}), \eprint{1111.2036}.

\bibitem[{\citenamefont{{Tang} and {Chevalier}}(2012)}]{2012Tang_apj}
\bibinfo{author}{\bibfnamefont{X.}~\bibnamefont{{Tang}}} \bibnamefont{and}
  \bibinfo{author}{\bibfnamefont{R.~A.} \bibnamefont{{Chevalier}}},
  \bibinfo{journal}{\apj} \textbf{\bibinfo{volume}{752}}, \bibinfo{eid}{83}
  (\bibinfo{year}{2012}), \eprint{1204.3913}.

\bibitem[{\citenamefont{{Torres} et~al.}(2014)\citenamefont{{Torres}, {Cillis},
  {Mart{\'\i}n}, and {de O{\~n}a Wilhelmi}}}]{2014Torres}
\bibinfo{author}{\bibfnamefont{D.~F.} \bibnamefont{{Torres}}},
  \bibinfo{author}{\bibfnamefont{A.}~\bibnamefont{{Cillis}}},
  \bibinfo{author}{\bibfnamefont{J.}~\bibnamefont{{Mart{\'\i}n}}},
  \bibnamefont{and} \bibinfo{author}{\bibfnamefont{E.}~\bibnamefont{{de O{\~n}a
  Wilhelmi}}}, \bibinfo{journal}{Journal of High Energy Astrophysics}
  \textbf{\bibinfo{volume}{1}}, \bibinfo{pages}{31} (\bibinfo{year}{2014}),
  \eprint{1402.5485}.

\bibitem[{\citenamefont{{Khangulyan} et~al.}(2020)\citenamefont{{Khangulyan},
  {Arakawa}, and {Aharonian}}}]{2020Khangulyan}
\bibinfo{author}{\bibfnamefont{D.}~\bibnamefont{{Khangulyan}}},
  \bibinfo{author}{\bibfnamefont{M.}~\bibnamefont{{Arakawa}}},
  \bibnamefont{and}
  \bibinfo{author}{\bibfnamefont{F.}~\bibnamefont{{Aharonian}}},
  \bibinfo{journal}{\mnras} \textbf{\bibinfo{volume}{491}},
  \bibinfo{pages}{3217} (\bibinfo{year}{2020}), \eprint{1911.07438}.

\bibitem[{\citenamefont{{Liang} et~al.}(2022)\citenamefont{{Liang}, {Li}, {Wu},
  {Pan}, and {Liu}}}]{2022Liang}
\bibinfo{author}{\bibfnamefont{X.-H.} \bibnamefont{{Liang}}},
  \bibinfo{author}{\bibfnamefont{C.-M.} \bibnamefont{{Li}}},
  \bibinfo{author}{\bibfnamefont{Q.-Z.} \bibnamefont{{Wu}}},
  \bibinfo{author}{\bibfnamefont{J.-S.} \bibnamefont{{Pan}}}, \bibnamefont{and}
  \bibinfo{author}{\bibfnamefont{R.-Y.} \bibnamefont{{Liu}}},
  \bibinfo{journal}{arXiv e-prints} \bibinfo{eid}{arXiv:2209.03809}
  (\bibinfo{year}{2022}), \eprint{2209.03809}.

\bibitem[{\citenamefont{{Lipari} and {Vernetto}}(2018)}]{2018Lipari}
\bibinfo{author}{\bibfnamefont{P.}~\bibnamefont{{Lipari}}} \bibnamefont{and}
  \bibinfo{author}{\bibfnamefont{S.}~\bibnamefont{{Vernetto}}},
  \bibinfo{journal}{\prd} \textbf{\bibinfo{volume}{98}}, \bibinfo{eid}{043003}
  (\bibinfo{year}{2018}), \eprint{1804.10116}.

\bibitem[{\citenamefont{{Di Mauro} et~al.}(2019)\citenamefont{{Di Mauro},
  {Manconi}, and {Donato}}}]{2019DiMauro}
\bibinfo{author}{\bibfnamefont{M.}~\bibnamefont{{Di Mauro}}},
  \bibinfo{author}{\bibfnamefont{S.}~\bibnamefont{{Manconi}}},
  \bibnamefont{and} \bibinfo{author}{\bibfnamefont{F.}~\bibnamefont{{Donato}}},
  \bibinfo{journal}{\prd} \textbf{\bibinfo{volume}{100}}, \bibinfo{eid}{123015}
  (\bibinfo{year}{2019}), \eprint{1903.05647}.

\bibitem[{\citenamefont{{Tang} and {Piran}}(2019)}]{2019Tang}
\bibinfo{author}{\bibfnamefont{X.}~\bibnamefont{{Tang}}} \bibnamefont{and}
  \bibinfo{author}{\bibfnamefont{T.}~\bibnamefont{{Piran}}},
  \bibinfo{journal}{\mnras} \textbf{\bibinfo{volume}{484}},
  \bibinfo{pages}{3491} (\bibinfo{year}{2019}), \eprint{1808.02445}.

\bibitem[{\citenamefont{{de Jager} and {Harding}}(1992)}]{1992dejager}
\bibinfo{author}{\bibfnamefont{O.~C.} \bibnamefont{{de Jager}}}
  \bibnamefont{and} \bibinfo{author}{\bibfnamefont{A.~K.}
  \bibnamefont{{Harding}}}, \bibinfo{journal}{\apj}
  \textbf{\bibinfo{volume}{396}}, \bibinfo{pages}{161} (\bibinfo{year}{1992}).

\bibitem[{\citenamefont{{Amato} and {Olmi}}(2021)}]{2021Amato}
\bibinfo{author}{\bibfnamefont{E.}~\bibnamefont{{Amato}}} \bibnamefont{and}
  \bibinfo{author}{\bibfnamefont{B.}~\bibnamefont{{Olmi}}},
  \bibinfo{journal}{Universe} \textbf{\bibinfo{volume}{7}},
  \bibinfo{pages}{448} (\bibinfo{year}{2021}), \eprint{2111.07712}.

\bibitem[{\citenamefont{{de O{\~n}a Wilhelmi} et~al.}(2022)\citenamefont{{de
  O{\~n}a Wilhelmi}, {L{\'o}pez-Coto}, {Amato}, and {Aharonian}}}]{2022emma}
\bibinfo{author}{\bibfnamefont{E.}~\bibnamefont{{de O{\~n}a Wilhelmi}}},
  \bibinfo{author}{\bibfnamefont{R.}~\bibnamefont{{L{\'o}pez-Coto}}},
  \bibinfo{author}{\bibfnamefont{E.}~\bibnamefont{{Amato}}}, \bibnamefont{and}
  \bibinfo{author}{\bibfnamefont{F.}~\bibnamefont{{Aharonian}}},
  \bibinfo{journal}{\apjl} \textbf{\bibinfo{volume}{930}}, \bibinfo{eid}{L2}
  (\bibinfo{year}{2022}), \eprint{2204.09440}.

\bibitem[{\citenamefont{{J{\'o}hannesson}
  et~al.}(2019)\citenamefont{{J{\'o}hannesson}, {Porter}, and
  {Moskalenko}}}]{2019Johannesson}
\bibinfo{author}{\bibfnamefont{G.}~\bibnamefont{{J{\'o}hannesson}}},
  \bibinfo{author}{\bibfnamefont{T.~A.} \bibnamefont{{Porter}}},
  \bibnamefont{and} \bibinfo{author}{\bibfnamefont{I.~V.}
  \bibnamefont{{Moskalenko}}}, \bibinfo{journal}{\apj}
  \textbf{\bibinfo{volume}{879}}, \bibinfo{eid}{91} (\bibinfo{year}{2019}),
  \eprint{1903.05509}.

\bibitem[{\citenamefont{{Martin} et~al.}(2022)\citenamefont{{Martin},
  {Marcowith}, and {Tibaldo}}}]{2022Martin}
\bibinfo{author}{\bibfnamefont{P.}~\bibnamefont{{Martin}}},
  \bibinfo{author}{\bibfnamefont{A.}~\bibnamefont{{Marcowith}}},
  \bibnamefont{and}
  \bibinfo{author}{\bibfnamefont{L.}~\bibnamefont{{Tibaldo}}},
  \bibinfo{journal}{arXiv e-prints} \bibinfo{eid}{arXiv:2206.11803}
  (\bibinfo{year}{2022}), \eprint{2206.11803}.

\bibitem[{\citenamefont{{Saz Parkinson} et~al.}(2010)\citenamefont{{Saz
  Parkinson}, {Dormody}, {Ziegler}, {Ray}, {Abdo}, {Ballet}, {Baring},
  {Belfiore}, {Burnett}, {Caliandro} et~al.}}]{SazParkinson2010}
\bibinfo{author}{\bibfnamefont{P.~M.} \bibnamefont{{Saz Parkinson}}},
  \bibinfo{author}{\bibfnamefont{M.}~\bibnamefont{{Dormody}}},
  \bibinfo{author}{\bibfnamefont{M.}~\bibnamefont{{Ziegler}}},
  \bibinfo{author}{\bibfnamefont{P.~S.} \bibnamefont{{Ray}}},
  \bibinfo{author}{\bibfnamefont{A.~A.} \bibnamefont{{Abdo}}},
  \bibinfo{author}{\bibfnamefont{J.}~\bibnamefont{{Ballet}}},
  \bibinfo{author}{\bibfnamefont{M.~G.} \bibnamefont{{Baring}}},
  \bibinfo{author}{\bibfnamefont{A.}~\bibnamefont{{Belfiore}}},
  \bibinfo{author}{\bibfnamefont{T.~H.} \bibnamefont{{Burnett}}},
  \bibinfo{author}{\bibfnamefont{G.~A.} \bibnamefont{{Caliandro}}},
  \bibnamefont{et~al.}, \bibinfo{journal}{\apj} \textbf{\bibinfo{volume}{725}},
  \bibinfo{pages}{571} (\bibinfo{year}{2010}), \eprint{1006.2134}.

\bibitem[{\citenamefont{{Yusifov} and
  {K{\"u}{\c{c}}{\"u}k}}(2004)}]{2004Yusifov}
\bibinfo{author}{\bibfnamefont{I.}~\bibnamefont{{Yusifov}}} \bibnamefont{and}
  \bibinfo{author}{\bibfnamefont{I.}~\bibnamefont{{K{\"u}{\c{c}}{\"u}k}}},
  \bibinfo{journal}{\aap} \textbf{\bibinfo{volume}{422}}, \bibinfo{pages}{545}
  (\bibinfo{year}{2004}), \eprint{astro-ph/0405559}.

\bibitem[{\citenamefont{{Acero} et~al.}(2016)\citenamefont{{Acero},
  {Ackermann}, {Ajello}, {Albert}, {Baldini}, {Ballet}, {Barbiellini},
  {Bastieri}, {Bellazzini}, {Bissaldi} et~al.}}]{Fermi2016}
\bibinfo{author}{\bibfnamefont{F.}~\bibnamefont{{Acero}}},
  \bibinfo{author}{\bibfnamefont{M.}~\bibnamefont{{Ackermann}}},
  \bibinfo{author}{\bibfnamefont{M.}~\bibnamefont{{Ajello}}},
  \bibinfo{author}{\bibfnamefont{A.}~\bibnamefont{{Albert}}},
  \bibinfo{author}{\bibfnamefont{L.}~\bibnamefont{{Baldini}}},
  \bibinfo{author}{\bibfnamefont{J.}~\bibnamefont{{Ballet}}},
  \bibinfo{author}{\bibfnamefont{G.}~\bibnamefont{{Barbiellini}}},
  \bibinfo{author}{\bibfnamefont{D.}~\bibnamefont{{Bastieri}}},
  \bibinfo{author}{\bibfnamefont{R.}~\bibnamefont{{Bellazzini}}},
  \bibinfo{author}{\bibfnamefont{E.}~\bibnamefont{{Bissaldi}}},
  \bibnamefont{et~al.}, \bibinfo{journal}{\apjs}
  \textbf{\bibinfo{volume}{223}}, \bibinfo{eid}{26} (\bibinfo{year}{2016}),
  \eprint{1602.07246}.

\bibitem[{\citenamefont{{Yang} et~al.}(2016)\citenamefont{{Yang}, {Aharonian},
  and {Evoli}}}]{Yang2016}
\bibinfo{author}{\bibfnamefont{R.}~\bibnamefont{{Yang}}},
  \bibinfo{author}{\bibfnamefont{F.}~\bibnamefont{{Aharonian}}},
  \bibnamefont{and} \bibinfo{author}{\bibfnamefont{C.}~\bibnamefont{{Evoli}}},
  \bibinfo{journal}{\prd} \textbf{\bibinfo{volume}{93}}, \bibinfo{eid}{123007}
  (\bibinfo{year}{2016}), \eprint{1602.04710}.

\bibitem[{\citenamefont{{Fang} et~al.}(2019)\citenamefont{{Fang}, {Bi}, and
  {Yin}}}]{Fang19}
\bibinfo{author}{\bibfnamefont{K.}~\bibnamefont{{Fang}}},
  \bibinfo{author}{\bibfnamefont{X.-J.} \bibnamefont{{Bi}}}, \bibnamefont{and}
  \bibinfo{author}{\bibfnamefont{P.-F.} \bibnamefont{{Yin}}},
  \bibinfo{journal}{\apj} \textbf{\bibinfo{volume}{884}}, \bibinfo{eid}{124}
  (\bibinfo{year}{2019}), \eprint{1906.08542}.

\bibitem[{\citenamefont{{Manconi} et~al.}(2020)\citenamefont{{Manconi}, {Di
  Mauro}, and {Donato}}}]{Manconi20}
\bibinfo{author}{\bibfnamefont{S.}~\bibnamefont{{Manconi}}},
  \bibinfo{author}{\bibfnamefont{M.}~\bibnamefont{{Di Mauro}}},
  \bibnamefont{and} \bibinfo{author}{\bibfnamefont{F.}~\bibnamefont{{Donato}}},
  \bibinfo{journal}{\prd} \textbf{\bibinfo{volume}{102}}, \bibinfo{eid}{023015}
  (\bibinfo{year}{2020}), \eprint{2001.09985}.

\end{thebibliography}

\clearpage
\appendix
\section{Fraction of the Halo Flux with an Unsuppressed Diffusion Coefficient}\label{app:a}
\begin{figure*}[htbp]
%\hspace{0.9cm} 
\centering
\includegraphics[width=0.45\textwidth]{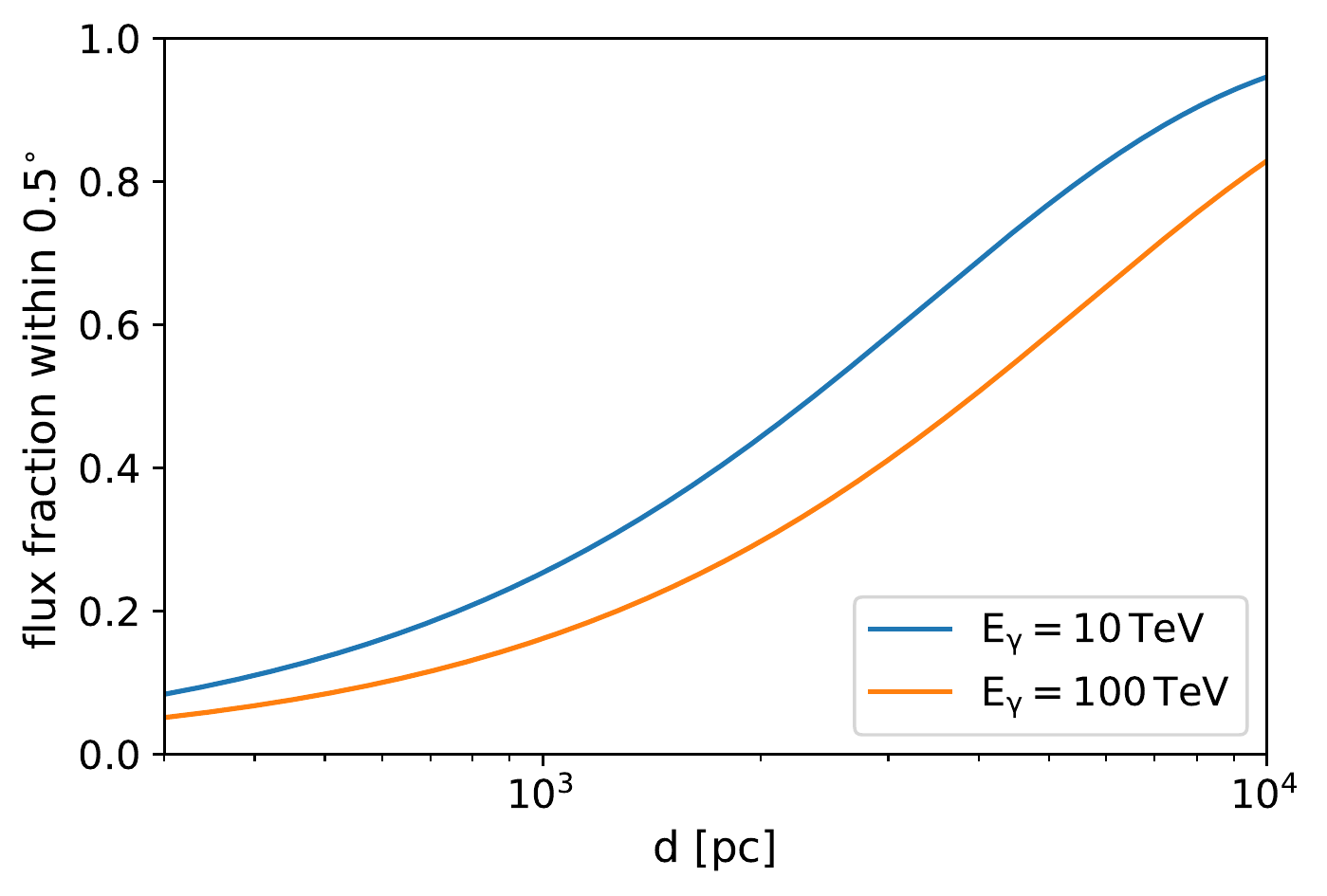}
\includegraphics[width=0.45\textwidth]{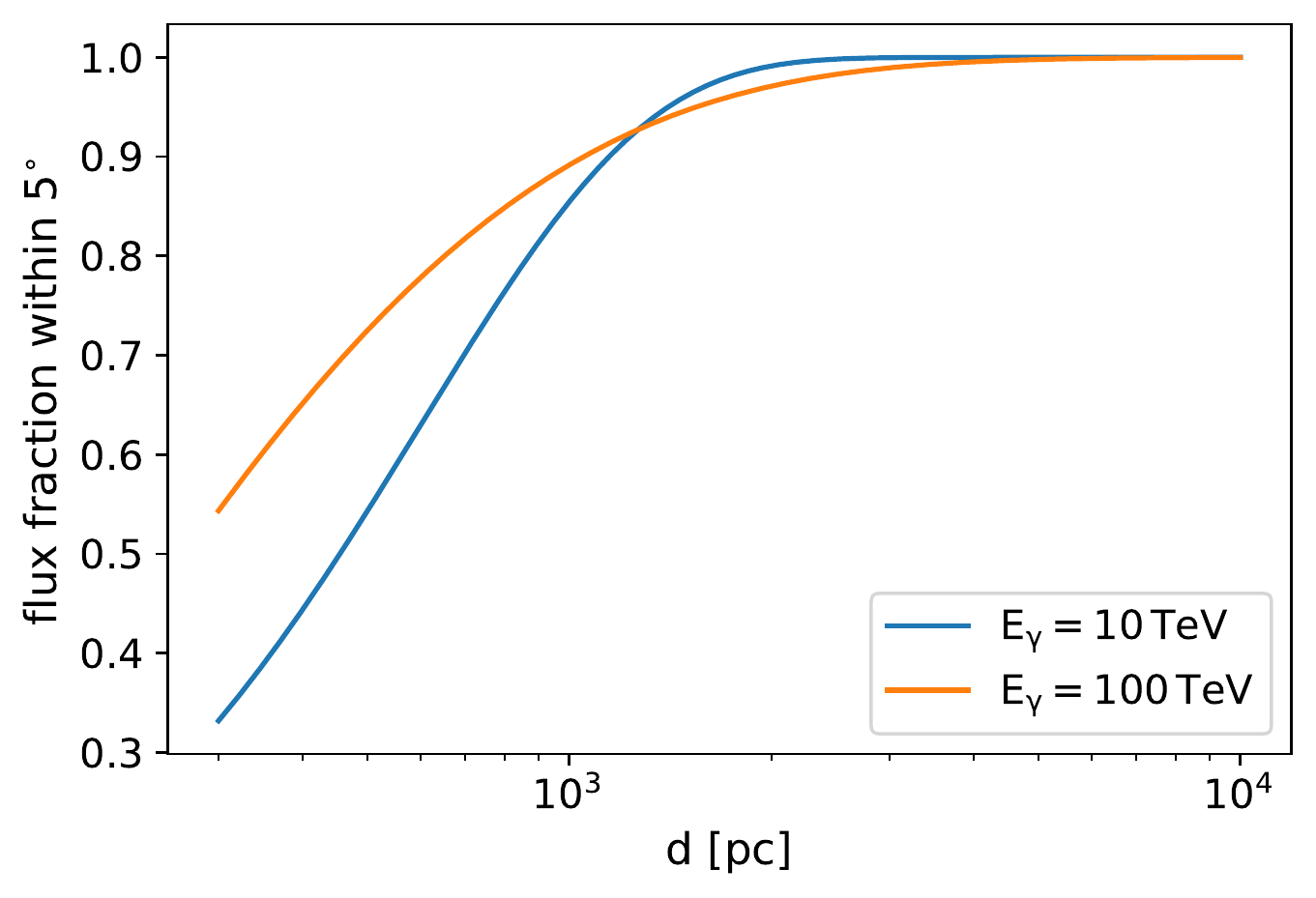}
\caption{Left: Flux fraction of a pulsar halo within $0.5^\circ$ from the pulsar as a function of the pulsar's distance. The diffusion coefficient is assumed to be $4 \times 10^{28}(E/1{\rm \,GeV})^{1/3}  \, \rm cm^2/s$. The magnetic field and the radiation field are the same as mentioned in Section~\ref{sec:Sec2.2}. Right: Same as the left panel but for the flux fraction within $5^\circ$ from the pulsar.}
\label{fig:append2}
\end{figure*}

When considering the ballistic propagation of particles at small radii (i.e., $r<3D/c$, see Ref.\citep{Recchia21}) with the standard ISM diffusion coefficient, i.e., $D(E)=4\times 10^{28}(E/1{\rm GeV})^{1/3}\rm cm^2/s$, the halo size would become quite extended. We calculate the surface brightness profile of a pulsar halo in the unsuppressed diffusion scenario following Ref.\citep{Recchia21}. In Fig.\,\ref{fig:append2} we plot the fraction of the halo flux contained within $0.5^\circ$ (left) and $5^\circ$ (right) from a pulsar, at 10\,TeV (blue curve) and 100\,TeV (orange curve), as a function of the pulsar's distance from Earth in this scenario. It can be seen that more than 50\% of the halo's emission at 10\,TeV (100\,TeV) would be beyond $0.5^\circ$ if the pulsar is located within 2\,kpc (4\,kpc) from Earth. Therefore, simply removing the pulsar within $0.5^\circ$ of any known TeV sources from the sample may lead to an underestimation of the DGE under the unsuppressed diffusion scenario by a factor of a few.
On the other hand, less than 10\% of the halo's emission would be distributed beyond $5^\circ$ from the pulsar as long as the pulsar is farther than 1\,kpc from Earth. In our pulsar sample, 4 of them are located within 1\,kpc and more than 80\% of the total emission of pulsar halo population are contained within $5^\circ$ from the pulsars. Therefore, it is safe to assume most of the energies of injected pairs are deposited in the Galactic plane, regardless of the transport model of pulsar halos.  %For a pulsar located at 1\,kpc, more than 90\% of the TeV emission will be contained within 5 degrees away from the source. That is to say, most of the TeV emission from pulsar halos in the Galactic disk will be contained within the region of Galactic latitude of plus or minus 5 degrees, even under the unsuppressed diffusion scenario, as we discussed in \ref{sec:Intro}. While on the other hand, more than half of the emission will extend to 0.5 degrees away, for a pulsar at 1\,kpc. So the mask region of 0.5 degrees would not be enough and nearby halos within the mask region could partially contribute to the DGE. In this sense, we have underestimated the DGE contributed by the pulsar halos and obtained conservative constraints on the injection parameters as discussed in Sec.\,\ref{sec:sec4b}.

\section{Pulsar Samples}\label{app:b}
Currently, the number of pulsars with characteristic age between 100\,kyr and 10\,Myr with known distance (including those inferred from dispersion measure) is 1164. Among them, 296 pulsars are within the ROI for AS$\rm \gamma$ (i.e., $25^{\circ}<l<100^{\circ}$, $\mid b\mid < 5^{\circ}$) and 142 samples are within the ROI for MILAGRO (i.e., $40^{\circ}<l<100^{\circ}$,$\mid b\mid < 5^{\circ}$). After performing the mask procedure and the removal of PSR J1952+3252 as introduced in Sec~\ref{sec:Sec2.1}, 236 pulsar samples remained as potential sources contributing to the observed DGE for AS$\rm \gamma$, and 217 pulsar samples remained for MILAGRO. The detailed information of these pulsars is listed in Table.~\ref{tab:pulsar}

\begin{figure*}[htbp]
%\hspace{0.9cm} 
\centering
\includegraphics[width=0.45\textwidth]{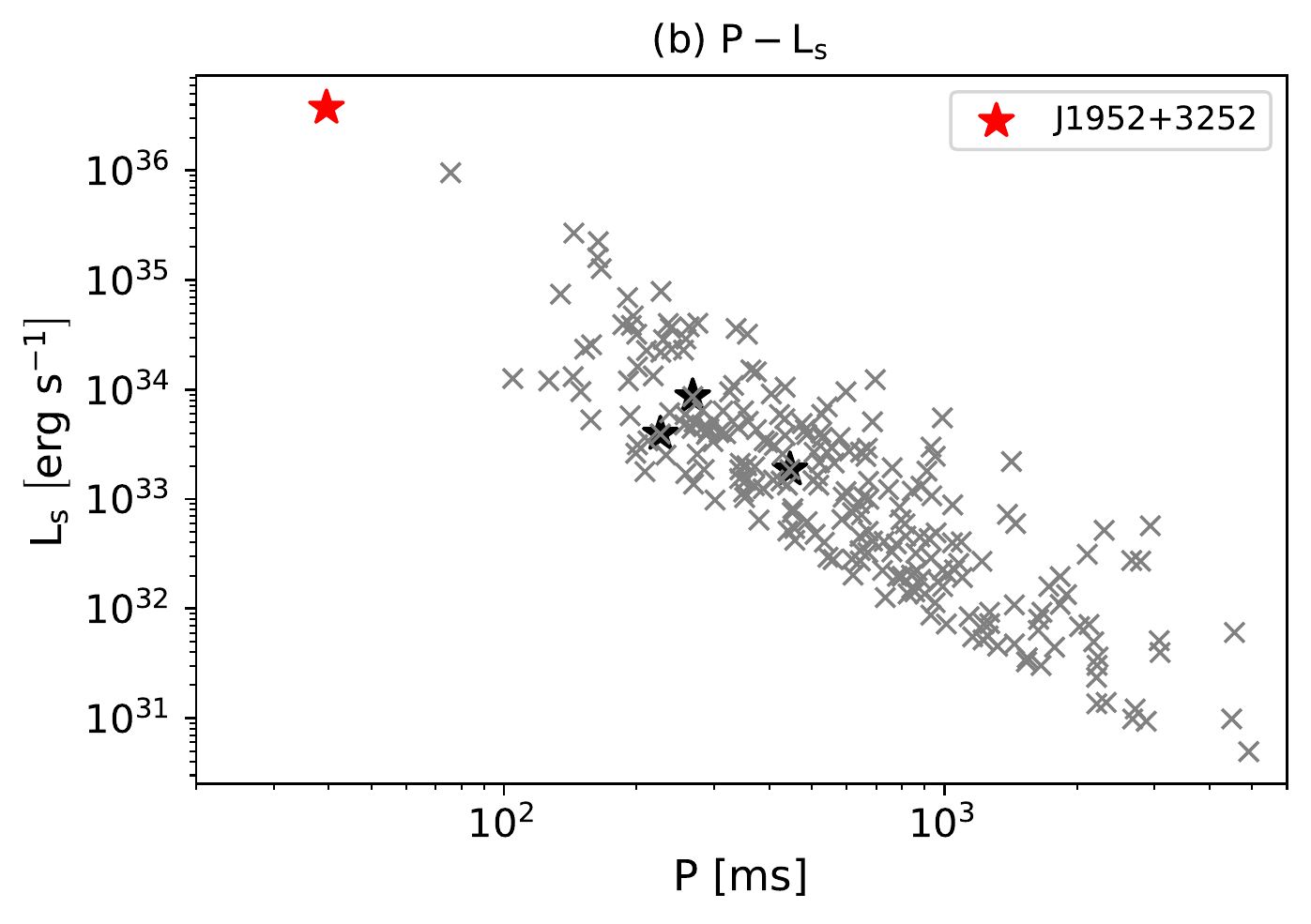}
\includegraphics[width=0.45\textwidth]{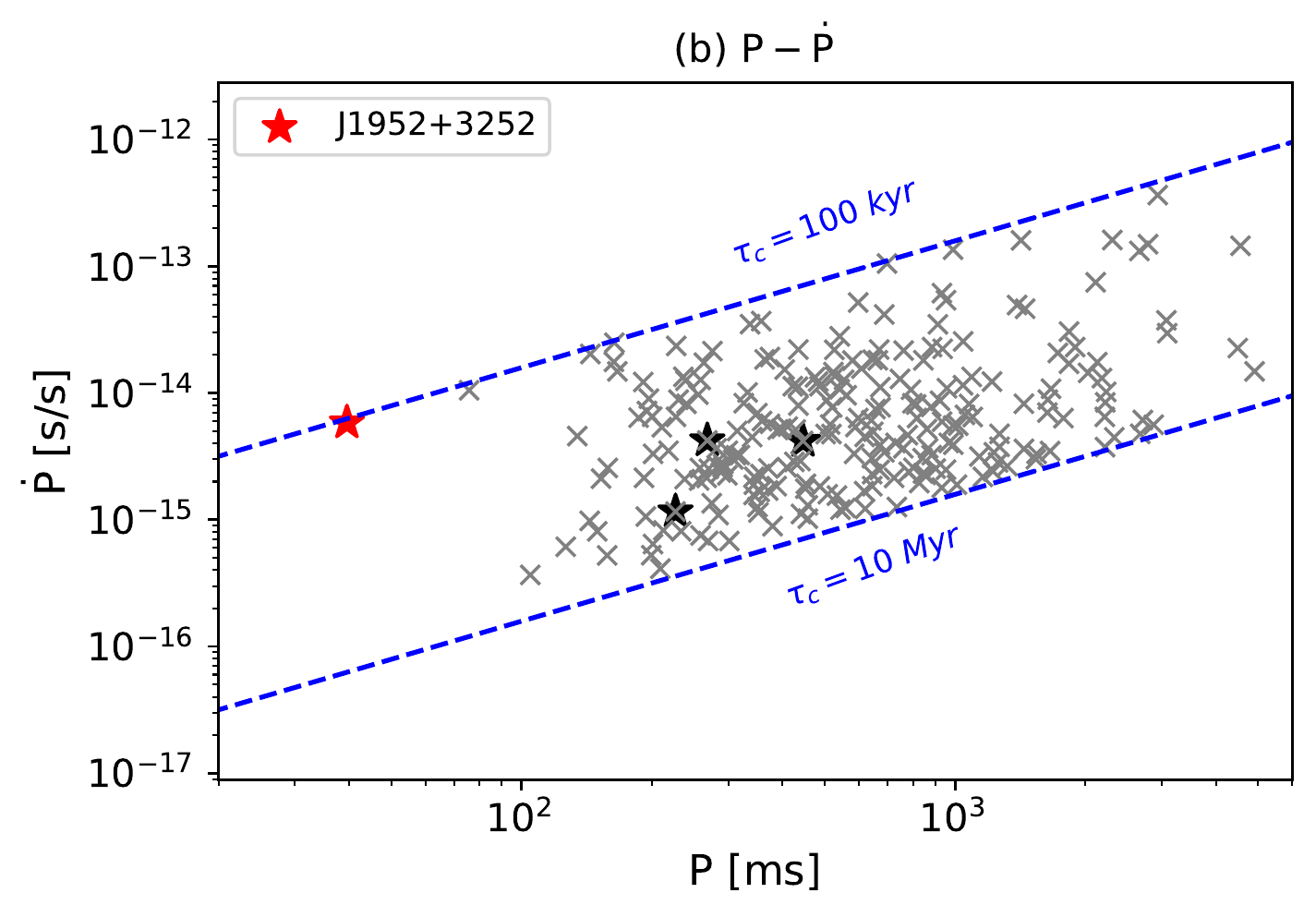}
\caption{Left: Distributions of pulsars in the sample (Table.~\ref{tab:pulsar}) including PSR~J1952+3252 in the $P -- L_s$ plane. Stars (crosses) represent pulsars with (without) SNR association, where PSR~J1952+3252 is marked as the red star; Right: Same as the left panel but for the distribution in $P--\dot{P}$ plane. The two dashed blue lines indicate the characteristic age of pulsars at 100\,kyr and 10\,Myr respectively.}
\label{fig:append1}
\end{figure*}

Among all the samples, PSR J1952+3252 outstands for its high spin-down luminosity, short rotation period, and association with an SNR, as shown in Fig.~\ref{fig:append1}. Other three pulsars of SNR associations have relatively slow rotation and low spin-down luminosity. The pulsar with the second highest spin-down luminosity and the second shortest period, PSR~J1925+1720, does not have an SNR association. Because of the specificity, we speculate that PSR J1952+3252 might be much younger than its characteristic age and exclude it in our calculation.

\clearpage
%\centering
\begin{center}
\begin{longtable}[b]{lllllll}
\caption{Middle-aged pulsar samples selected in our calculation within $25^{\circ}<l<100^{\circ}$, $\mid b\mid < 5^{\circ}$}\\
\hline
\hline
    JName & $l \left[^{\circ}\right]$ & $b \left[^{\circ}\right]$& $P \left[ \rm s\right] $& $\dot{P} \left[ \rm s/s \right]$ & $L_{\rm s} \left[ \rm \rm erg \ s^{-1}\right]$ & $d \left[ \rm \rm kpc\right]$\\
\hline
    J1820-0427 & 25.46 & 4.73 & 0.598082 & 6.33e-15 & 1.17e+33 & 2.857 \\
    J1821-0331 & 26.39 & 4.98 & 0.902316 & 2.53e-15 & 1.36e+32 & 7.556 \\
    J1829+0000 & 30.46 & 4.82 & 0.199147 & 5.25e-16 & 2.62e+33 & 4.353 \\
    J1830-0131 & 29.16 & 3.99 & 0.152512 & 2.11e-15 & 2.34e+34 & 3.502 \\
    J1832+0029 & 31.25 & 4.36 & 0.533918 & 1.55e-15 & 4.03e+32 & 1.120 \\
    J1833-0209 & 28.92 & 3.09 & 0.291931 & 2.75e-15 & 4.37e+33 & 13.360 \\
    J1833-0338 & 27.66 & 2.27 & 0.686733 & 4.16e-14 & 5.07e+33 & 2.500 \\
    J1833-0559 & 25.51 & 1.32 & 0.483459 & 1.23e-14 & 4.31e+33 & 6.827 \\
    J1834-0602 & 25.64 & 0.97 & 0.487914 & 1.83e-15 & 6.21e+32 & 6.340 \\
    J1835-0349 & 27.68 & 1.86 & 0.841865 & 3.06e-15 & 2.02e+32 & 5.510 \\
    J1835-0600 & 25.76 & 0.83 & 2.221787 & 8.43e-15 & 3.03e+31 & 10.644 \\
    J1836-0436 & 27.17 & 1.13 & 0.354237 & 1.66e-15 & 1.48e+33 & 4.358 \\
    J1836-0517 & 26.51 & 0.92 & 0.457245 & 1.30e-15 & 5.38e+32 & 8.315 \\
    J1837-0045 & 30.67 & 2.75 & 0.617037 & 1.68e-15 & 2.83e+32 & 3.145 \\
    J1837-0559 & 26.00 & 0.38 & 0.201064 & 3.31e-15 & 1.61e+34 & 4.315 \\
    J1839-0141 & 30.01 & 1.97 & 0.933266 & 5.94e-15 & 2.89e+32 & 6.074 \\
    J1839-0223 & 29.50 & 1.46 & 1.26679 & 4.76e-15 & 9.25e+31 & 6.088 \\
    J1839-0321 & 28.60 & 1.10 & 0.238782 & 1.25e-14 & 3.63e+34 & 7.852 \\
    J1839-0332 & 28.46 & 0.93 & 2.675682 & 4.76e-15 & 9.81e+30 & 4.042 \\
    J1839-0402 & 28.02 & 0.73 & 0.52094 & 7.69e-15 & 2.15e+33 & 4.231 \\
    J1839-0436 & 27.41 & 0.65 & 0.149461 & 8.1e-16 & 9.57e+33 & 4.483 \\
    J1839-0459 & 27.15 & 0.32 & 0.585319 & 3.31e-15 & 6.51e+32 & 3.945 \\
    J1840+0214 & 33.70 & 3.44 & 0.797478 & 8.29e-15 & 6.46e+32 & 5.846 \\
    J1840-0445 & 27.49 & 0.20 & 0.422316 & 1.13e-14 & 5.91e+33 & 4.557 \\
    J1840-0626 & 25.93 & -0.46 & 1.893353 & 2.30e-14 & 1.34e+32 & 5.727 \\
    J1841-0157 & 30.10 & 1.22 & 0.663321 & 1.81e-14 & 2.45e+33 & 7.930 \\
    J1841-0425 & 27.82 & 0.28 & 0.186149 & 6.39e-15 & 3.91e+34 & 4.399 \\
    J1841-0500 & 27.32 & -0.03 & 0.912916 & 3.48e-14 & 1.80e+33 & 4.969 \\
    J1842-0153 & 30.28 & 1.02 & 1.054228 & 6.72e-15 & 2.26e+32 & 6.875 \\
    J1842+0257 & 34.56 & 3.34 & 3.088256 & 2.96e-14 & 3.97e+31 & 4.990 \\
    J1842+0358 & 35.43 & 3.85 & 0.233326 & 8.11e-16 & 2.52e+33 & 4.329 \\
    J1842-0415 & 28.09 & 0.11 & 0.526682 & 2.19e-14 & 5.93e+33 & 3.605 \\
    J1843-0000 & 32.01 & 1.77 & 0.880334 & 7.77e-15 & 4.50e+32 & 3.336 \\
    J1843+0119 & 33.20 & 2.39 & 1.266998 & 3.76e-15 & 7.29e+31 & 6.069 \\
    J1843-0137 & 30.54 & 1.09 & 0.669872 & 2.47e-15 & 3.24e+32 & 7.662 \\
    J1843-0211 & 30.08 & 0.77 & 2.027524 & 1.44e-14 & 6.84e+31 & 5.931 \\
    J1843-0408 & 28.37 & -0.17 & 0.781934 & 2.39e-15 & 1.97e+32 & 3.983 \\
    J1843-0510 & 27.39 & -0.52 & 0.671614 & 3.89e-15 & 5.07e+32 & 4.068 \\
    J1843-0702 & 25.74 & -1.43 & 0.191615 & 2.14e-15 & 1.20e+34 & 4.285 \\
    J1843-0744 & 25.09 & -1.68 & 0.475393 & 1.33e-14 & 4.89e+33 & 7.053 \\
    J1844-0030 & 31.71 & 1.27 & 0.641098 & 6.08e-15 & 9.11e+32 & 10.411 \\
    J1844-0433 & 28.10 & -0.55 & 0.991027 & 3.91e-15 & 1.59e+32 & 3.074 \\
    J1844-0452 & 27.75 & -0.58 & 0.269443 & 6.80e-16 & 1.37e+33 & 5.684 \\
    J1844-0538 & 27.07 & -0.94 & 0.255704 & 9.71e-15 & 2.29e+34 & 5.404 \\
    J1845-0434 & 28.19 & -0.78 & 0.486751 & 1.13e-14 & 3.88e+33 & 4.093 \\
    J1845-0545 & 27.15 & -1.34 & 1.092348 & 1.34e-14 & 4.07e+32 & 5.471 \\
    J1845-0635 & 26.35 & -1.60 & 0.340528 & 4.49e-15 & 4.49e+33 & 9.407 \\
    J1845-0743 & 25.43 & -2.30 & 0.104695 & 3.67e-16 & 1.26e+34 & 7.113 \\
    J1846+0051 & 33.16 & 1.44 & 0.434373 & 1.12e-14 & 5.41e+33 & 3.998 \\
    J1846-0749 & 25.39 & -2.43 & 0.35011 & 1.26e-15 & 1.16e+33 & 13.990 \\
    J1846-07492 & 25.37 & -2.39 & 0.86138 & 5.19e-15 & 3.20e+32 & 4.220 \\
    J1847-0402 & 28.88 & -0.94 & 0.597809 & 5.17e-14 & 9.55e+33 & 3.419 \\
    J1847-0438 & 28.37 & -1.27 & 0.957991 & 1.09e-14 & 4.91e+32 & 4.378 \\
    J1847-0605 & 27.05 & -1.87 & 0.778164 & 4.65e-15 & 3.89e+32 & 4.324 \\
    J1848-0055 & 31.80 & 0.17 & 0.274557 & 1.35e-15 & 2.57e+33 & 7.408 \\
    J1848-0511 & 27.94 & -1.66 & 1.637129 & 8.86e-15 & 7.97e+31 & 8.799 \\
    J1848+0604 & 38.06 & 3.33 & 2.218603 & 3.74e-15 & 1.35e+31 & 12.600 \\
    J1848+0647 & 38.70 & 3.65 & 0.505957 & 8.75e-15 & 2.67e+33 & 1.128 \\
    J1849-0040 & 32.08 & 0.20 & 0.672481 & 1.11e-14 & 1.45e+33 & 7.887 \\
    J1849+0106 & 33.74 & 0.84 & 1.832259 & 1.70e-14 & 1.09e+32 & 4.581 \\
    \hline 
\hline
\hline
    JName & $l \left[^{\circ}\right]$ & $b \left[^{\circ}\right]$& $P \left[ \rm s\right] $& $\dot{P} \left[ \rm s/s \right]$ & $L_{\rm s} \left[ \rm \rm erg \ s^{-1}\right]$ & $d \left[ \rm \rm kpc\right]$\\
    \hline
    J1849+0127 & 34.03 & 1.04 & 0.542155 & 2.80e-14 & 6.93e+33 & 4.691 \\
    J1849-0317 & 29.83 & -1.17 & 0.668408 & 2.20e-14 & 2.91e+33 & 1.212 \\
    J1849+0409 & 36.37 & 2.42 & 0.761194 & 2.16e-14 & 1.93e+33 & 1.974 \\
    J1849-0614 & 27.18 & -2.47 & 0.953384 & 5.39e-14 & 2.46e+33 & 3.504 \\
    J1849-0636 & 26.77 & -2.50 & 1.451319 & 4.62e-14 & 5.97e+32 & 3.849 \\
    J1850-0031 & 32.37 & -0.04 & 0.734185 & 1.26e-15 & 1.26e+32 & 6.502 \\
    J1851-0029 & 32.54 & -0.34 & 0.518721 & 4.74e-15 & 1.34e+33 & 5.453 \\
    J1851-0114 & 31.81 & -0.53 & 0.953182 & 2.48e-15 & 1.13e+32 & 5.214 \\
    J1851+0233 & 35.18 & 1.23 & 0.344018 & 2.18e-15 & 2.12e+33 & 12.653 \\
    J1851+0241 & 35.31 & 1.25 & 4.491318 & 2.26e-14 & 9.83e+30 & 11.117 \\
    J1851-0241 & 30.52 & -1.19 & 0.435194 & 7.96e-15 & 3.81e+33 & 7.922 \\
    J1851+0418 & 36.72 & 2.05 & 0.284697 & 1.09e-15 & 1.86e+33 & 4.089 \\
    J1852-0118 & 31.87 & -0.78 & 0.451473 & 1.76e-15 & 7.54e+32 & 4.726 \\
    J1852-0127 & 31.71 & -0.80 & 0.428979 & 5.15e-15 & 2.57e+33 & 5.692 \\
    J1852-0635 & 27.22 & -3.34 & 0.524157 & 1.46e-14 & 4.01e+33 & 4.577 \\
    J1853+0056 & 34.02 & -0.04 & 0.275578 & 2.14e-14 & 4.03e+34 & 3.841 \\
    J1853+0427 & 37.17 & 1.51 & 1.320659 & 2.65e-15 & 4.53e+31 & 15.719 \\
    J1853+0545 & 38.35 & 2.06 & 0.1264 & 6.12e-16 & 1.20e+34 & 6.519 \\
    J1854+0306 & 35.99 & 0.83 & 4.55782 & 1.45e-13 & 6.05e+31 & 4.498 \\
    J1854-0524 & 28.51 & -3.24 & 0.544021 & 1.20e-15 & 2.94e+32 & 5.014 \\
    J1855+0306 & 36.17 & 0.48 & 1.633566 & 7.00e-15 & 6.34e+31 & 7.422 \\
    J1855+0307 & 36.17 & 0.53 & 0.845348 & 1.81e-14 & 1.18e+33 & 5.940 \\
    J1855+0700 & 39.61 & 2.34 & 0.258685 & 7.52e-16 & 1.71e+33 & 10.336 \\
    J1856+0102 & 34.43 & -0.65 & 0.620217 & 1.22e-15 & 2.02e+32 & 6.540 \\
    J1856-0526 & 28.64 & -3.58 & 0.370483 & 1.70e-15 & 1.32e+33 & 4.034 \\
    J1857+0526 & 38.44 & 1.19 & 0.349951 & 6.93e-15 & 6.38e+33 & 12.231 \\
    J1858+0346 & 37.08 & 0.18 & 0.256844 & 2.04e-15 & 4.75e+33 & 5.490 \\
    J1859+0601 & 39.24 & 0.90 & 1.044313 & 2.55e-14 & 8.84e+32 & 6.364 \\
    J1859+0603 & 39.27 & 0.93 & 0.508561 & 1.59e-15 & 4.77e+32 & 9.003 \\
    J1900-0134 & 32.55 & -2.72 & 1.832332 & 3.05e-14 & 1.96e+32 & 4.854 \\
    J1900+0227 & 36.17 & -0.93 & 0.374262 & 5.71e-15 & 4.30e+33 & 4.515 \\
    J1900+0438 & 38.07 & 0.17 & 0.312314 & 3.23e-15 & 4.19e+33 & 6.988 \\
    J1900+0634 & 39.81 & 0.99 & 0.389869 & 5.13e-15 & 3.41e+33 & 8.595 \\
    J1901+0124 & 35.38 & -1.68 & 0.318817 & 3.24e-15 & 3.95e+33 & 6.597 \\
    J1901+0156 & 35.82 & -1.37 & 0.288219 & 2.36e-15 & 3.89e+33 & 3.229 \\
    J1901+0234 & 36.37 & -1.05 & 0.88524 & 2.30e-14 & 1.31e+33 & 7.523 \\
    J1901-0312 & 31.19 & -3.65 & 0.355725 & 2.29e-15 & 2.01e+33 & 3.778 \\
    J1901-0315 & 31.15 & -3.67 & 0.801693 & 2.57e-15 & 1.97e+32 & 8.445 \\
    J1901+0331 & 37.21 & -0.64 & 0.65545 & 7.46e-15 & 1.05e+33 & 7.000 \\
    J1901+0355 & 37.58 & -0.44 & 0.554756 & 1.27e-14 & 2.95e+33 & 6.685 \\
    J1901+0413 & 37.81 & -0.23 & 2.66308 & 1.32e-13 & 2.75e+32 & 5.342 \\
    J1902+0248 & 36.74 & -1.25 & 1.223777 & 2.41e-15 & 5.18e+31 & 5.990 \\
    J1902+0556 & 39.50 & 0.21 & 0.746577 & 1.29e-14 & 1.22e+33 & 3.600 \\
    J1902+0615 & 39.81 & 0.34 & 0.673505 & 7.71e-15 & 9.96e+32 & 7.000 \\
    J1903+0135 & 35.73 & -1.96 & 0.729307 & 4.03e-15 & 4.10e+32 & 3.300 \\
    J1903-0258 & 31.66 & -4.04 & 0.301459 & 6.79e-16 & 9.79e+32 & 4.069 \\
    J1903+0601 & 39.65 & 0.11 & 0.374117 & 1.92e-14 & 1.45e+34 & 5.887 \\
    J1904-0150 & 32.83 & -3.84 & 0.379387 & 8.90e-16 & 6.43e+32 & 5.305 \\
    J1905-0056 & 33.69 & -3.55 & 0.643183 & 3.06e-15 & 4.54e+32 & 7.644 \\
    J1905+0600 & 39.84 & -0.28 & 0.44121 & 1.11e-15 & 5.11e+32 & 8.797 \\
    J1906+0414 & 38.48 & -1.51 & 1.043362 & 1.15e-14 & 3.98e+32 & 10.069 \\
    J1906+0509 & 39.29 & -1.08 & 0.39759 & 5.22e-15 & 3.28e+33 & 3.091 \\
    J1907+0249 & 37.31 & -2.32 & 0.351879 & 1.14e-15 & 1.03e+33 & 8.751 \\
    J1907+0345 & 38.08 & -1.80 & 0.240153 & 8.22e-15 & 2.34e+34 & 9.481 \\
    J1907+0534 & 39.72 & -0.99 & 1.138403 & 3.15e-15 & 8.43e+31 & 11.835 \\
    J1908+0500 & 39.29 & -1.40 & 0.291021 & 2.59e-15 & 4.14e+33 & 5.845 \\
    J1909+0007 & 35.12 & -3.98 & 1.016948 & 5.52e-15 & 2.07e+32 & 4.358 \\
    J1909+0254 & 37.60 & -2.71 & 0.989831 & 5.53e-15 & 2.25e+32 & 4.500 \\
    J1910+0358 & 38.61 & -2.34 & 2.330263 & 4.47e-15 & 1.39e+31 & 2.859 \\
    J1914+0219 & 37.63 & -4.04 & 0.457527 & 1.02e-15 & 4.20e+32 & 14.386 \\
    J1857+0809 & 40.84 & 2.45 & 0.502924 & 4.74e-15 & 1.47e+33 & 13.138 \\
\hline
\hline
    JName & $l \left[^{\circ}\right]$ & $b \left[^{\circ}\right]$& $P \left[ \rm s\right] $& $\dot{P} \left[ \rm s/s \right]$ & $L_{\rm s} \left[ \rm \rm erg \ s^{-1}\right]$ & $d \left[ \rm \rm kpc\right]$\\
    \hline
    J1901+0716 & 40.57 & 1.06 & 0.643999 & 2.29e-15 & 3.38e+32 & 3.400 \\
    J1902+1141 & 44.54 & 2.98 & 0.40914 & 2.59e-15 & 1.49e+33 & 13.888 \\
    J1903+0654 & 40.50 & 0.39 & 0.791232 & 1.06e-14 & 8.44e+32 & 5.923 \\
    J1903+0912 & 42.52 & 1.49 & 0.166314 & 1.48e-14 & 1.27e+35 & 11.845 \\
    J1903+0925 & 42.74 & 1.54 & 0.357155 & 3.69e-14 & 3.20e+34 & 6.256 \\
    J1904+0738 & 41.18 & 0.68 & 0.208958 & 4.11e-16 & 1.78e+33 & 6.154 \\
    J1904+0800 & 41.50 & 0.86 & 0.263345 & 1.73e-14 & 3.74e+34 & 10.966 \\
    J1905+0616 & 40.07 & -0.17 & 0.989706 & 1.35e-13 & 5.51e+33 & 4.952 \\
    J1905+0709 & 40.94 & 0.06 & 0.64804 & 4.94e-15 & 7.17e+32 & 4.980 \\
    J1905+0902 & 42.56 & 1.06 & 0.218253 & 3.5e-15 & 1.33e+34 & 11.553 \\
    J1905+1034 & 43.92 & 1.76 & 1.72681 & 2.07e-14 & 1.59e+32 & 6.910 \\
    J1906+0641 & 40.60 & -0.30 & 0.267275 & 2.14e-15 & 4.42e+33 & 7.000 \\
    J1906+0724 & 41.22 & 0.07 & 1.53649 & 3.00e-15 & 3.26e+31 & 6.929 \\
    J1906+0746 & 41.60 & 0.15 & 0.144073 & 2.03e-14 & 2.68e+35 & 7.400 \\
    J1907+0731 & 41.50 & -0.21 & 0.363676 & 1.84e-14 & 1.51e+34 & 4.994 \\
    J1907+1149 & 45.29 & 1.83 & 1.42016 & 1.60e-13 & 2.20e+33 & 7.551 \\
    J1907+1247 & 46.10 & 2.37 & 0.827097 & 1.95e-15 & 1.36e+32 & 10.526 \\
    J1908+0734 & 41.59 & -0.27 & 0.212353 & 8.25e-16 & 3.40e+33 & 0.669 \\
    J1908+0909 & 42.97 & 0.49 & 0.336555 & 3.49e-14 & 3.61e+34 & 8.906 \\
    J1909+1102 & 44.83 & 0.99 & 0.283642 & 2.64e-15 & 4.57e+33 & 4.800 \\
    J1909+1205 & 45.78 & 1.47 & 1.229312 & 3.40e-15 & 7.23e+31 & 10.130 \\
    J1910+0534 & 40.06 & -1.67 & 0.452867 & 1.92e-15 & 8.18e+32 & 21.255 \\
    J1910+0714 & 41.52 & -0.87 & 2.712423 & 6.12e-15 & 1.21e+31 & 3.679 \\
    J1910+0728 & 41.74 & -0.77 & 0.325415 & 8.31e-15 & 9.52e+33 & 6.233 \\
    J1910+1017 & 44.25 & 0.52 & 0.411159 & 5.42e-15 & 3.08e+33 & 13.677 \\
    J1910+1231 & 46.21 & 1.59 & 1.441742 & 8.23e-15 & 1.08e+32 & 8.139 \\
    J1911+1051 & 44.89 & 0.50 & 0.190873 & 1.22e-14 & 6.91e+34 & 10.118 \\
    J1911+1301 & 46.79 & 1.54 & 1.010462 & 1.89e-15 & 7.23e+31 & 11.653 \\
    J1912+2104 & 54.09 & 4.99 & 2.232969 & 1.02e-14 & 3.61e+31 & 3.369 \\
    J1913+0657 & 41.64 & -1.71 & 1.257181 & 2.83e-15 & 5.62e+31 & 5.069 \\
    J1913+0832 & 42.98 & -0.86 & 0.134409 & 4.57e-15 & 7.43e+34 & 8.204 \\
    J1913+0904 & 43.50 & -0.68 & 0.163246 & 1.76e-14 & 1.60e+35 & 2.997 \\
    J1913+1330 & 47.42 & 1.38 & 0.923391 & 8.68e-15 & 4.35e+32 & 6.179 \\
    J1914+1428 & 48.46 & 1.49 & 1.15952 & 2.18e-15 & 5.52e+31 & 6.514 \\
    J1915+0639 & 41.66 & -2.37 & 0.64414 & 1.84e-15 & 2.72e+32 & 8.874 \\
    J1915+0738 & 42.47 & -1.8 & 1.542704 & 3.31e-15 & 3.55e+31 & 1.404 \\
    J1915+0838 & 43.34 & -1.3 & 0.342777 & 1.57e-15 & 1.54e+33 & 10.840 \\
    J1915+1009 & 44.71 & -0.65 & 0.404552 & 1.53e-14 & 9.10e+33 & 7.000 \\
    J1916+0748 & 42.77 & -2.05 & 0.541752 & 1.07e-14 & 2.66e+33 & 11.779 \\
    J1916+0844 & 43.54 & -1.49 & 0.439995 & 2.90e-15 & 1.34e+33 & 10.988 \\
    J1916+0852 & 43.67 & -1.45 & 2.182746 & 1.31e-14 & 4.97e+31 & 9.391 \\
    J1916+0951 & 44.56 & -1.02 & 0.270254 & 2.52e-15 & 5.04e+33 & 1.904 \\
    J1916+1225 & 46.81 & 0.23 & 0.227387 & 2.35e-14 & 7.87e+34 & 6.486 \\
    J1916+1312 & 47.58 & 0.45 & 0.281845 & 3.66e-15 & 6.45e+33 & 4.500 \\
    J1917+0834 & 43.58 & -1.89 & 2.129665 & 1.75e-14 & 7.15e+31 & 1.231 \\
    J1917+1353 & 48.26 & 0.62 & 0.194631 & 7.20e-15 & 3.85e+34 & 5.882 \\
    J1917+2224 & 55.78 & 4.55 & 0.425897 & 2.86e-15 & 1.46e+33 & 4.965 \\
    J1918+1311 & 47.76 & 0.06 & 0.856749 & 2.26e-15 & 1.42e+32 & 6.187 \\
    J1918+1541 & 49.89 & 1.36 & 0.370883 & 2.54e-15 & 1.97e+33 & 0.727 \\
    J1920+1040 & 45.78 & -1.59 & 2.215802 & 6.48e-15 & 2.35e+31 & 10.151 \\
    J1921+0812 & 43.71 & -2.93 & 0.210648 & 5.36e-15 & 2.27e+34 & 2.896 \\
    J1921+0921 & 44.73 & -2.42 & 0.562302 & 9.58e-15 & 2.13e+33 & 6.139 \\
    J1921+1544 & 50.35 & 0.61 & 0.143576 & 9.80e-16 & 1.31e+34 & 9.038 \\
    J1921+1630 & 50.95 & 1.14 & 0.936448 & 2.23e-14 & 1.07e+33 & 5.097 \\
    J1922+1733 & 52.08 & 1.23 & 0.236171 & 1.34e-14 & 4.01e+34 & 5.360 \\
    J1922+2110 & 55.28 & 2.94 & 1.077924 & 8.18e-15 & 2.58e+32 & 4.000 \\
    J1924+1631 & 51.40 & 0.32 & 2.935186 & 3.64e-13 & 5.69e+32 & 10.183 \\
    J1924+1639 & 51.42 & 0.56 & 0.158043 & 2.56e-15 & 2.56e+34 & 5.059 \\
    J1924+2040 & 55.02 & 2.33 & 0.23779 & 2.09e-15 & 6.14e+33 & 5.951 \\
    J1925+1720 & 52.18 & 0.59 & 0.075659 & 1.05e-14 & 9.54e+35 & 5.048 \\
    J1926+1648 & 51.86 & 0.06 & 0.579823 & 1.8e-14 & 3.64e+33 & 6.000 \\
\hline
\hline
    JName & $l \left[^{\circ}\right]$ & $b \left[^{\circ}\right]$& $P \left[ \rm s\right] $& $\dot{P} \left[ \rm s/s \right]$ & $L_{\rm s} \left[ \rm \rm erg \ s^{-1}\right]$ & $d \left[ \rm \rm kpc\right]$\\
    \hline
    J1926+2016 & 54.85 & 1.80 & 0.299072 & 3.50e-15 & 5.17e+33 & 5.948 \\
    J1927+1856 & 53.81 & 0.94 & 0.298313 & 2.24e-15 & 3.34e+33 & 3.100 \\
    J1928+1923 & 54.28 & 1.02 & 0.81733 & 6.35e-15 & 4.59e+32 & 10.583 \\
    J1929+1357 & 49.63 & -1.81 & 0.866927 & 3.66e-15 & 2.22e+32 & 4.768 \\
    J1929+1955 & 54.88 & 1.02 & 0.257832 & 2.56e-15 & 5.89e+33 & 6.603 \\
    J1929+2121 & 56.12 & 1.75 & 0.723599 & 2.14e-15 & 2.23e+32 & 2.653 \\
    J1930+1316 & 49.12 & -2.32 & 0.760032 & 3.66e-15 & 3.29e+32 & 6.339 \\
    J1931+1439 & 50.54 & -2.01 & 1.779226 & 6.33e-15 & 4.44e+31 & 6.123 \\
    J1931+1536 & 51.41 & -1.60 & 0.314355 & 5.01e-15 & 6.37e+33 & 4.011 \\
    J1932+1059 & 47.38 & -3.88 & 0.226519 & 1.16e-15 & 3.93e+33 & 0.310 \\
    J1932+2020 & 55.58 & 0.64 & 0.268217 & 4.22e-15 & 8.63e+33 & 5.000 \\
    J1933+2421 & 59.48 & 2.39 & 0.81369 & 8.11e-15 & 5.94e+32 & 4.639 \\
    J1935+1616 & 52.44 & -2.09 & 0.358738 & 6.00e-15 & 5.13e+33 & 3.700 \\
    J1935+1829 & 54.36 & -1.0 & 0.843548 & 2.32e-15 & 1.53e+32 & 8.616 \\
    J1936+1536 & 51.88 & -2.46 & 0.967338 & 4.04e-15 & 1.76e+32 & 4.624 \\
    J1936+2042 & 56.38 & -0.07 & 1.390723 & 4.94e-14 & 7.25e+32 & 4.999 \\
    J1937+1505 & 51.57 & -2.98 & 2.872774 & 5.61e-15 & 9.34e+30 & 6.450 \\
    J1937+2544 & 60.84 & 2.27 & 0.20098 & 6.43e-16 & 3.13e+33 & 3.125 \\
    J1937+2950 & 64.50 & 4.12 & 1.657429 & 3.48e-15 & 3.02e+31 & 7.477 \\
    J1938+2010 & 56.12 & -0.67 & 0.687082 & 3.40e-15 & 4.14e+32 & 8.812 \\
    J1938+2659 & 62.11 & 2.56 & 0.883332 & 3.23e-15 & 1.85e+32 & 8.576 \\
    J1939+2449 & 60.17 & 1.36 & 0.645302 & 1.83e-14 & 2.68e+33 & 7.115 \\
    J1940+2245 & 58.63 & 0.13 & 0.258912 & 1.27e-14 & 2.89e+34 & 8.081 \\
    J1941+1341 & 50.80 & -4.47 & 0.559084 & 1.24e-15 & 2.80e+32 & 5.459 \\
    J1941+2525 & 61.04 & 1.26 & 2.306153 & 1.61e-13 & 5.18e+32 & 11.168 \\
    J1946+2535 & 61.81 & 0.28 & 0.515167 & 5.64e-15 & 1.63e+33 & 8.305 \\
    J1946+2611 & 62.32 & 0.60 & 0.43506 & 2.2e-14 & 1.05e+34 & 7.555 \\
    J1947+1957 & 56.99 & -2.66 & 0.157509 & 5.23e-16 & 5.28e+33 & 6.767 \\
    J1948+2333 & 60.21 & -1.04 & 0.528352 & 1.36e-14 & 3.63e+33 & 8.048 \\
    J1948+2551 & 62.21 & 0.13 & 0.196627 & 9.02e-15 & 4.69e+34 & 8.698 \\
    J1948+2819 & 64.37 & 1.31 & 0.932693 & 6.13e-14 & 2.98e+33 & 11.042 \\
    J1950+3001 & 66.09 & 1.75 & 2.788918 & 1.49e-13 & 2.71e+32 & 8.759 \\
    J1952+3021 & 66.53 & 1.65 & 1.665665 & 1.08e-14 & 9.25e+31 & 7.485 \\
    J1954+2407 & 61.37 & -1.87 & 0.193405 & 1.06e-15 & 5.76e+33 & 4.176 \\
    J1958+3033 & 67.35 & 0.69 & 1.098581 & 6.46e-15 & 1.92e+32 & 7.266 \\
    J2000+2920 & 66.55 & -0.35 & 3.073783 & 3.74e-14 & 5.09e+31 & 6.684 \\
    J2002+3217 & 69.26 & 0.88 & 0.696761 & 1.05e-13 & 1.23e+34 & 6.458 \\
    J2004+3137 & 69.01 & 0.02 & 2.111265 & 7.46e-14 & 3.13e+32 & 8.000 \\
    J2005+3552 & 72.71 & 2.14 & 0.307943 & 2.99e-15 & 4.04e+33 & 14.652 \\
    J2006+3102 & 68.67 & -0.53 & 0.163695 & 2.49e-14 & 2.24e+35 & 6.035 \\
    J2007+3120 & 69.04 & -0.54 & 0.608205 & 1.56e-14 & 2.74e+33 & 6.906 \\
    J2008+2513 & 64.06 & -4.11 & 0.589196 & 5.40e-15 & 1.04e+33 & 4.030 \\
    J2010+3230 & 70.39 & -0.50 & 1.442448 & 3.62e-15 & 4.76e+31 & 13.034 \\
    J2011+3331 & 71.32 & -0.05 & 0.931733 & 1.79e-15 & 8.72e+31 & 8.603 \\
    J2013+3845 & 75.93 & 2.48 & 0.230194 & 8.85e-15 & 2.86e+34 & 7.123 \\
    J2018+3431 & 73.04 & -0.84 & 0.387664 & 1.84e-15 & 1.24e+33 & 6.636 \\
    J2022+2854 & 68.86 & -4.67 & 0.343402 & 1.89e-15 & 1.85e+33 & 2.100 \\
    J2029+3744 & 76.90 & -0.73 & 1.216805 & 1.23e-14 & 2.70e+32 & 5.771 \\
    J2030+3641 & 76.12 & -1.44 & 0.200129 & 6.50e-15 & 3.20e+34 & 6.947 \\
    J2030+4415 & 82.34 & 2.88 & 0.22707 & 6.48e-15 & 2.19e+34 & 0.720 \\
    J2037+3621 & 76.75 & -2.84 & 0.618715 & 4.50e-15 & 7.50e+32 & 4.851 \\
    J2047+5029 & 89.06 & 4.38 & 0.445945 & 4.18e-15 & 1.86e+33 & 3.973 \\
    J2053+4718 & 87.21 & 1.62 & 4.910379 & 1.48e-14 & 4.94e+30 & 8.901 \\
    J2150+5247 & 97.52 & -0.92 & 0.332206 & 1.01e-14 & 1.09e+34 & 3.610 \\
\hline
\label{tab:pulsar}
\end{longtable}
\end{center}

\section{Distance-dependent correction factor}\label{app:c}
\begin{figure*}[htbp]
%\hspace{0.9cm} 
\centering
\includegraphics[width=0.45\textwidth]{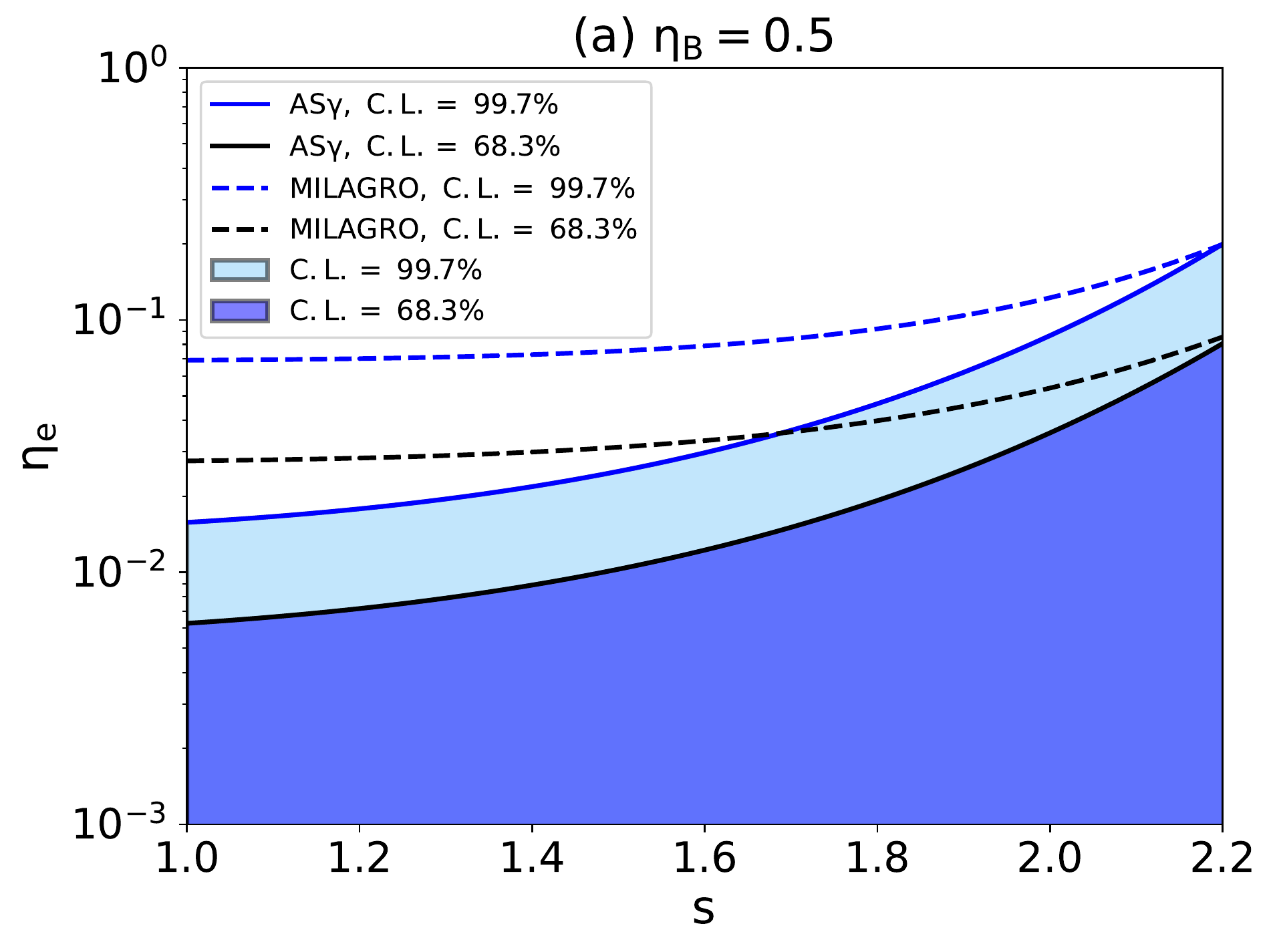}
\includegraphics[width=0.45\textwidth]{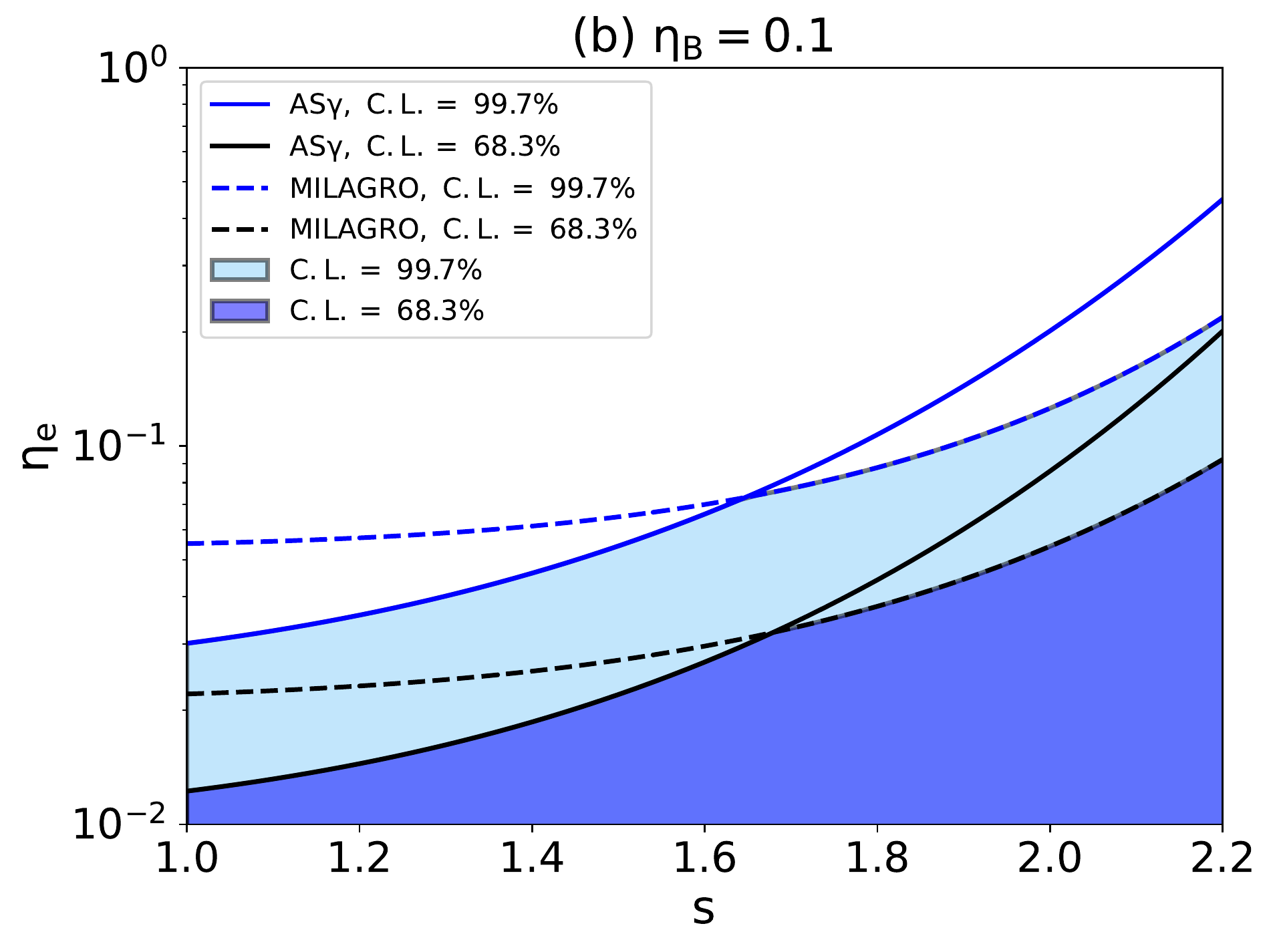}
\caption{Two-dimensional constraints between $s$ and $\eta_{\rm e}$ after considering selection effect of pulsar detection. Other details are the same as in Fig.~\ref{fig:etaB}.}
\label{fig:append}
\end{figure*}

%In the main text of this paper, we do not account for the gamma-ray emission from those pulsars with weak flux in the radio band. Here we follow \citet{2004Yusifov} to obtain a correction factor of the spatial density of pulsars and discuss the influence of those sub-threshold pulsars on our results. \citet{2004Yusifov} considered two types of selection effects. One is direction-dependent and connected to the variation of background radiation with Galactic longitudes.  
\citet{2004Yusifov} suggested two categories of instrumental selection effects: one is the direction-dependent (mostly longitudinal) effect which is related to the variation of noise from the sky, while the other is the distance-dependent effect because the flux of a pulsar is proportional to $1/d^2$. As the considered ROI ($25^{\circ}<l<100^{\circ}$) is away from the Galactic center, the direction-dependent factor does not vary much in the ROI. Thus, we only examine the influence of the distance-dependent selection effect. We follow \citet{2004Yusifov} to take the distance-dependent correction factor $f_{\rm d}=\rm exp\left(-c_0 d \right)$, where $c_0=0.362\pm 0.017$ and $d$ is the distance to the observer in unit of kpc. Then we weight the contribution of each single pulsar halo to the DGE by the correction factor $f_d$, as we did for the beaming factor $f_{\rm beam}$. Fig.~\ref{fig:append} shows the parameter constraints after considering the correction factor. The obtained upper limits of $\eta_e$ are reduced by about an order of magnitude after the correction. 
%We note, however, that the correction factor derived by \citet{2004Yusifov} is an overestimation for the pulsar sample at the present time due to the improvement of instruments sensitivity. To make the obtained conclusion of the paper on the safe side, we do not adopt the result with distance-dependent correction factor in the main text but just show it here for reference.

%\begin{figure*}[htbp]
%\hspace{0.9cm} 
%\includegraphics[width=1.0\textwidth]{Emax_without.pdf}\\
%\includegraphics[width=1.0\textwidth]{s_without.pdf}
%\caption{Same as Fig. 3, but without contributions from pp collisions.}
%\label{fig:Emax_without}
%\end{figure*}

%\nocite{*}

\end{document}